\documentclass[10pt,a4paper]{article}
\makeatletter
\selectfont

\renewcommand{\section}{\@startsection{section}{1}{\z@}%
{2ex}{1ex}{\reset@font\large\bfseries}}%
\renewcommand{\thesection}{\@arabic\c@section}
\def\@listi{\topsep=.3\baselineskip \parsep=.2ex \partopsep=0ex%
\itemsep=0ex \leftmargin=4ex \rightmargin=2ex}
\let\@listI\@listi
\@listi\def\@listii{\parsep=.2ex \partopsep=0pt \itemsep=0ex%
\leftmargin=4ex \rightmargin=0ex}
\let\@listiii\@listii
\let\@listiv\@listii
\let\@listv\@listii
\let\@listvi\@listii
\long\def\@makecaption#1#2{\footnotesize\sbox\@tempboxa{#1. #2}
\ifdim\wd\@tempboxa >\hsize #1. #2\par
\else \global\@minipagefalse
\hb@xt@\hsize{\hfil\box\@tempboxa\hfil}
\fi}
\makeatother
\usepackage{lineno}
\usepackage[pdftex,final]{graphicx} 
\usepackage[hyperindex]{hyperref}
\title{Mathematical modeling and analysis of the pathway network consisting of symmetrical complexes with N monomers, like the activation of MMP2.}
\author{Keiko Itano \thanks{Osaka University, itano@sigmath.es.osaka-u.ac.jp}}
\begin{document}

\maketitle

\begin{abstract}
 The activation of matrix metalloproteinase 2 (MMP2)  is a crucial event during tumor metastasis and invasion, and this pathway network consists of 3 monomers. The pathway network of the activation obeys to a set of specified reaction rules.
According to the rules, the individual molecules localize in a particular order and symmetrically around a homodimer following the formation of that dimer.

We generalized the homodimer pathway network obeying to similar reaction rules, by changing the number of monomers involved in this pathway from 3 to N. 
At the previous work, we found the molecules in the pathway network are classified to some reaction groups.
We derived the law of mass conservation between the groups. 
Each group concentration converges to its equilibrium solution.
Using these results, we derive the concentrations of the complexes theoretically and reveal that each complex concentration converges to its equilibrium value.
We can say the pathway network with homodimer symmetric form complexes is asymptotic stable and identify the regulator parameter of the target  complex in the network. 
Our mathematical approach may help us understand the mechanism of this type pathway network by knowing the background mathematical laws which govern this type pathway network.
\end{abstract}

%
\section{Introduction}
\begin{figure}
 \includegraphics[width=11cm]{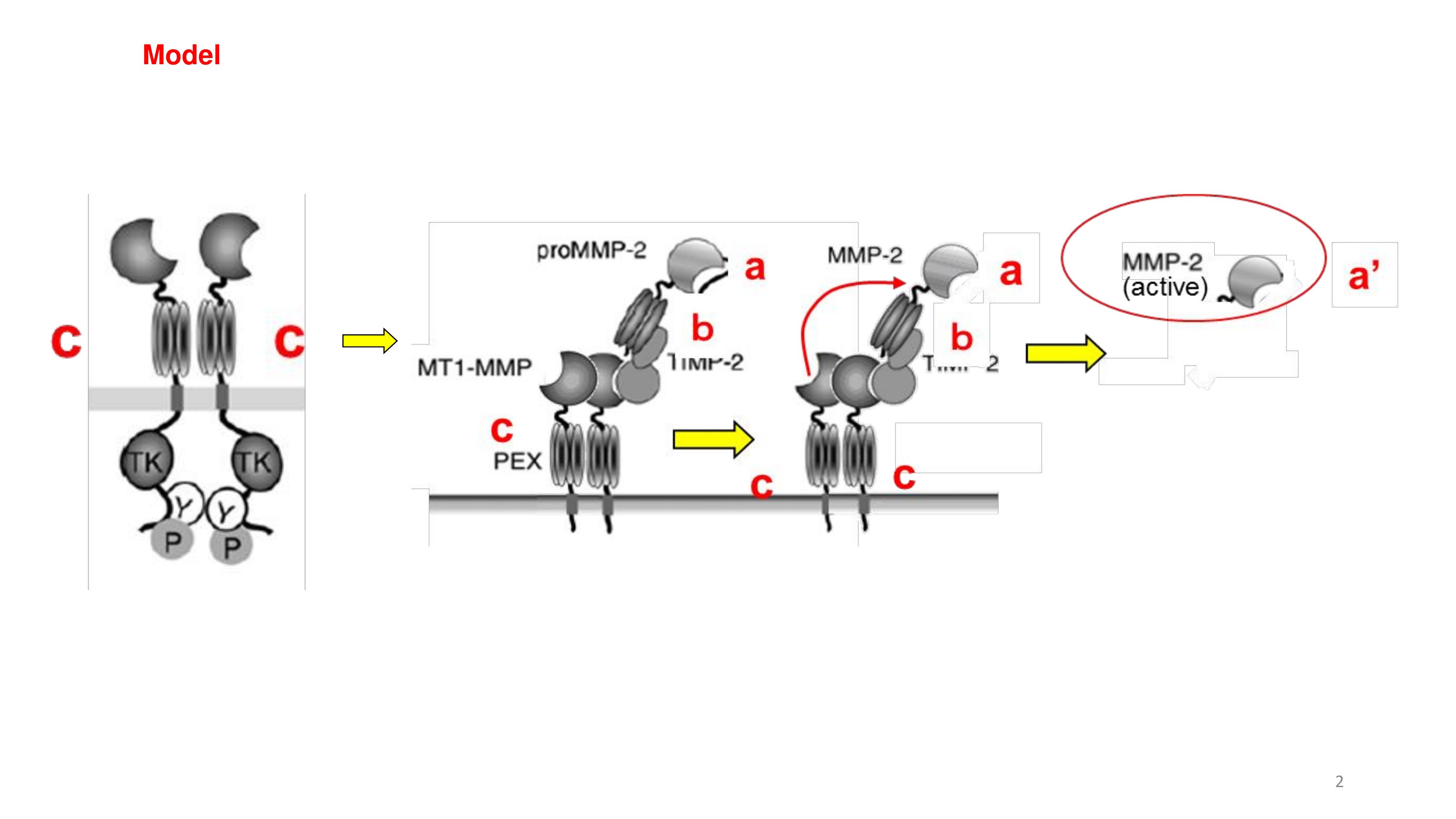}
  \caption{MT1-MMP complex activates MMP2.
      When the (MMP2 - TIMP2 - MT1-MMP - MT1-MMP) complex is formed, the interactions between MMP2 and TIMP2 are disrupted, which leads to the activation of MMP2.}
\label{fig:1}
\end{figure}

Matrix metalloproteinases (MMPs) represent a family of endopeptidases that are responsible for the degradation of many components of extracellular matrix (ECM). The process of ECM degradation plays an important role during tumor metastasis and invasion, and this pathway network consists of 3 monomers: matrix metalloproteinase 2 (MMP2), tissue inhibitor of metalloproteinase(TIMP2), and membrane type-1 matrix metalloproteinase(MT1-MMP). Sato \textit{et al.} (1994) demonstrated that MMP2 activation occurs upon the formation of (MMP2- TIMP2 - MT1-MMP - MT1-MMP) complex \cite{mtmmp} (Fig.\ref{fig:1}).
MMP inhibitors for cancer treatments have been developed, but their many side effects have hindered the use of these inhibitors. 
Therefore, the suppression of this complex is necessary in order to develop improved treatments, and the mathematical modeling mechanisms involved in the activation of MMP2 may represent a very useful approach in anti-cancer drug development.
Several mathematical approaches of the activation pathway of MMP 2 have been proposed.
Hoshino \textit{et al.}(2012) proposed a computational model of this (MMP2 - TIMP2- MT1-MMP - MT1-MMP) complex, taking into consideration the transient dynamics of the MMP2 activation \cite{hoshino}.
Saito \textit{et al.}(2012) investigated this computational model, and, using the computational simulations, found a novel drug target useful for the suppression of (MMP2-TIMP2- MT1-MMP - MT1-MMP) complex  during the MMP2 activation process \cite{saitou}.

The ordinary differential equations (ODEs), as the molecular network evolution equations, have too many terms to be solved theoretically. 
Kawasaki \textit{et al.} showed that the MMP2 activation pathway network can be divided into several groups, and that ODEs of the MMP2 activation network are solvable and all complexes have explicit solutions\cite{kawasaki}. 
Itano and Suzuki extended the 3-monomer model to the N monomer model and showed that the monomers in the extended pathway network can be classified into reaction groups. By grouping, the $N(N+1)$ molecule ODEs are summed into the $N$ group ODEs according to the law of mass conservation. The group ODEs are solved explicitely and have their equilibria. \cite{itano}

The aim of this study was to understand the mechanism of a biochemical reaction by generalizing a pathway network and solving the ODEs, in order to elucidate the relationships between the molecules involved in this pathway, to investigate the pathway network regulation, and to identify theoretically parameters important for the regulation of this network. The generalization of the pathway network, including N monomers, allows a better understanding of the biochemical reaction mechanism.

In the complexes formed during MMP2 activation, the monomers localize in a particular order, symmetrically around the MT1-MMP homodimer. We found that the formation of the molecular complexes in the networks depends on the chemical reaction rules of the network and decides 
 the pathway network evolution.

According to their role in the complex formation, the molecules of the network are classified into several reaction groups, with N mass conservation laws between the groups. Using these mass conservation laws, the N(N+1) molecule ODEs can be simplified to N group ODEs, which have strict solutions and the equilibrium solutions.
The equilibrium solutions of the molecule ODEs are calculated theoretically with the group solutions as their own upper values.

\begin{figure}
\includegraphics[ width=12cm]{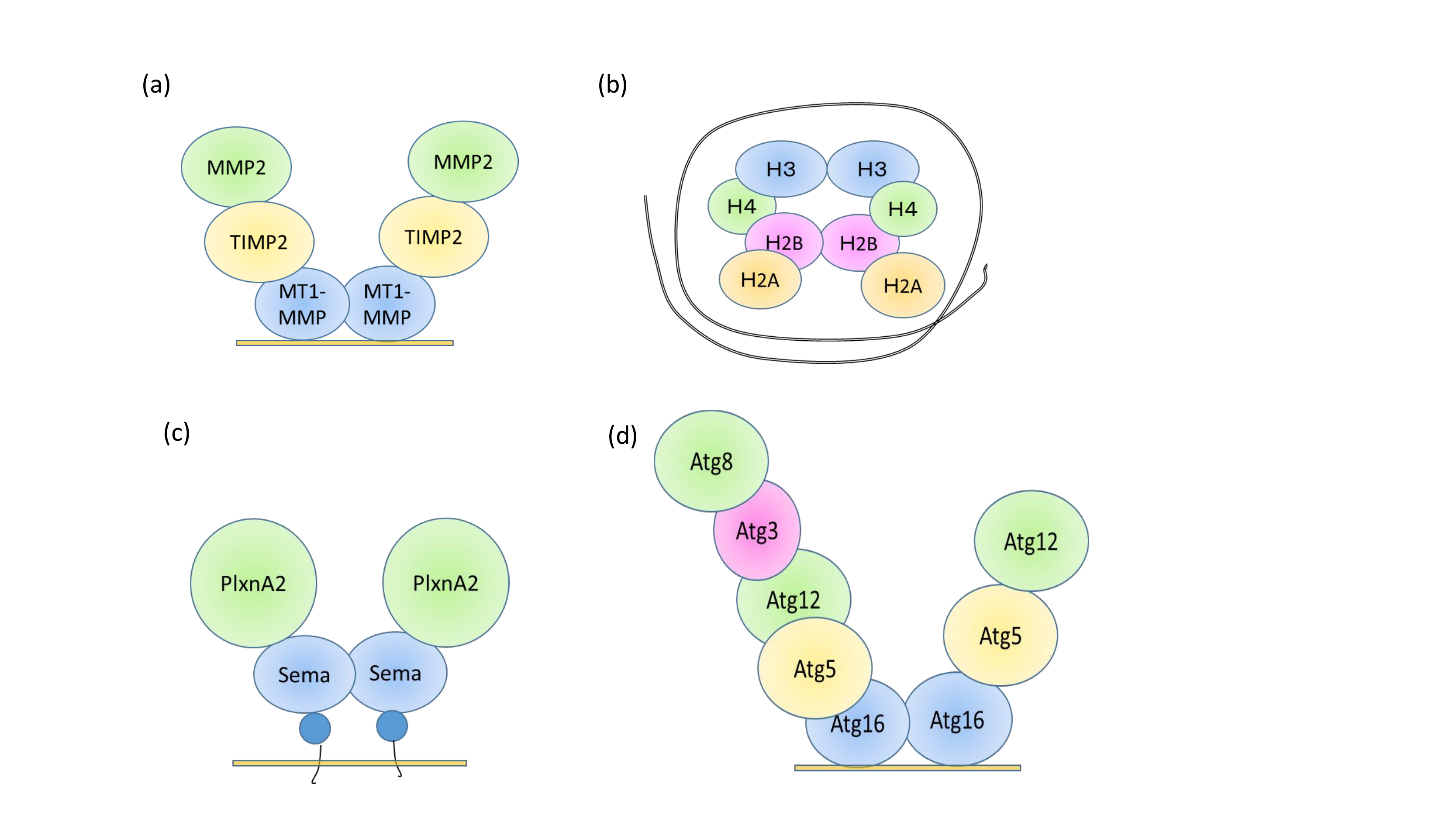}
  \caption{Examples of the symmetric formation of biological $(b_N-\cdots-b_1-b_1-\cdots-b_N)$ type complexes.
      (a) Histone octamer wrapped with DNA takes a (H2B-H2A-H4-H3-H3-H4-H2A-H2B) form. This complex plays an important role during transcription.
	(b) Semaphorin and plexin complex takes a (PlxnA2-Sema-Sema-PlxnA2) form. 
(c) (Atg12-Atg5-Atg16-Atg16-Atg5-Atg12) complex reacts with Atg3-Atg8 complex and forms (Atg8-Atg3-Atg12-Atg5-Atg16-Atg16-Atg5-Atg12). This complex formation is similar to that of (MMP2-TIMP2-MT1-MMP-MT1-MMP).}
\label{fig:2}
      \end{figure}

In this paper, we introduce MMP2 activation model as an example of a pathway network that shows a symmetrical complex formation,  ($b_N-\cdots-b_1-b_1-\cdots-b_N$) type complex formation. Most polymers form circular symmetric complexes, with the $b_N-b_N$ connections, because this ring formation allows for a more stable formation of the complexes. However, ($b_N-\cdots-b_1-b_1-\cdots-b_N$) formation type symmetric polymers exists, and it is possible to apply the results obtained in this study to the pathway networks in which molecules have similar symmetric complex formation, such as the formation of histone octamer \cite{histone}, autophagy complex (Atg8-Atg3-Atg12-Atg5-Atg16-Atg16-Atg5-Atg12)\cite{atgcomp} \cite{atgcomp2} \cite{atgcomp3}, semaphorin and plexin complex \cite{semapho}, Kai3-Kai2-Kai1 complex, immunoglobulin complex, and others (Fig.~\ref{fig:2})~ 

Shinar and  Feinberg investigated the robustness of network depending on its structure. They divided a pathway network into different modules, and showed the robustness of the network depends on the robustness of all reaction modules \cite{feinberg}.
Mochizuki and Fiedler describe the sensitivity of the flux response to a change in the reaction rate. They investigated the network structure in order to obtain the response sensitivity. For a pathway network with a layered structure, they divided the network to motifs and obtained a local flux response. Afterward, they obtained the global flux response sensitivity \cite{mochi}. In the network structure approach, a network is divided into several modules, and the local modules are independent of each other. The local module parameters or characteristic are obtained independently, while the total network characteristics are investigated globally.

Our approach is similar to the described approaches from the perspective that we classify the pathway network to some modules and clarify the structure. The extended N monomer pathway network has a layered structure 
because the complexes are classified into several groups according to their formation. We solved all molecular concentrations, using group solutions obtained previously \cite{itano}, where we solved group solutions locally. Here, we described and solved the complex solutions globally. 

\section{ N monomer model}
We generalized a 3-monomer pathway network, obtaining the N monomer pathway network. The reaction rules are similar to those in the 3-monomer MMP2 activation network. 
$b_1$ monomers react each other and make homodimer. Other monomers connects with monomers in a specified order. We show the reaction rules as follows.

\newpage
\begin{itemize}
 \item $b_1$ monomer reacts with $b_1$ monomer and makes a homodimer $b_1-b_1$. \\
$\hspace{1cm} b_1 + b_1 \rightarrow b_1b_1, $
\\($k_1$: reaction rate constant and $l_1$: the dissolution rate constant ) \\
 \item $b_i$ monomer reacts with $b_{i-1}$ or $b_{i+1}$ monomer. ($i=2, \cdots, N-1$).\\
$\hspace{1cm} b_{i+1} + b_{i} \rightarrow b_{i+1}b_i , $
\\( $k_{i+1}$:the reaction rate constant , and $l_{i+1}$:the dissolution rate constant) 
\end{itemize}
According to the reaction rules, the pathway network has $N(N+1)$ types of complexes as follows.
$\\b_N,  b_N b_{N-1}, \dots, b_N b_{N-1}\cdots b_2 b_1,  b_N\cdots b_1b_1, b_N\cdots b_1b_1 b_2, \dots, b_N\cdots b_1 b_1 \cdots b_N\\
b_{N-1},  b_{N-1} b_{N-2},  \dots,  b_{N-1}\cdots-b_1,  b_{N-1}\cdots b_1 b_1, \dots,  b_{N-1}\cdots b_1 b_1\cdots b_{N-1}\\
\hspace{1cm}  \vdots\\
b_2,  b_2 b_1, b_2 b_1b_1,\\
b_1,  b_1b_1 $
	
\subsection{$N$ monomer pathway network and its structural features}
The generalization helps us understand the structural features dependent on to the complex formations. We determined that different groups of complexes react with each other, in order to produce a new type of complex, and this grouping depends on the formation of complex and the type of the edge monomer. 

\subsection{Grouping and strict solutions of the groups}
The complexes belong to the reaction groups, and we determined mass conservation relationships between the reaction groups. $N(N+1)$ ODEs for complex evolution analyses were aggregated to the $N$ ODEs for reaction groups. The $N$ reaction group ODEs are independent of each other and quadratic with one variable, and therefore, solvable. We showed that reaction group ODEs have strict solutions and the solutions converge to the equilibrium state.

\subsection{Complex integrability}
The reaction group solutions were obtained. First, we obtained \textit{a priori} upper bound of the concentration of each complex in the group, and afterward, we show that each complex ODE is written as follows:
 \begin{eqnarray}
\frac{d X_{l,m}(t) }{dt} = -A_{l,m}(t)X_{l,m}(t)+f_{l,m}(t) , \hspace{1 cm}
\end{eqnarray}
where variables $A_{l,m}(t)$ and $f_{l,m}(t)$ can be written with reaction group variable $\xi_m(t)$ and reaction rate constants $k_m$ and $l_m$($m$=1, $\cdots$, N). Reaction group variable $\xi_m(t)$ is strictly derived.
The solution of this equation is:
\begin{eqnarray}
\label{SotionlCmpX}
 X_{l,m}(t) =  \mathrm{e}^{-\int_0^tA_{l,m}(s)ds}+\int_0^t  \mathrm{e}^{-\int_0^sA_{l,m}(u)du} f_{l,m}(s)ds. \hspace{1 cm}
\end{eqnarray}
This shows that each complex concentration has a solution, and the $N(N+1)$ complex ODEs are integrable.
Using the equation (\ref{SotionlCmpX}), we obtained the equilibrium solution. Therefore, the equilibrium solution of $X^{\ast}_{l, m} = \lim_{t \to \infty}X_{l, m}(t)$ is:
\begin{eqnarray}
\label{EqulibriumXn}
X^{\ast}_{l, m} =\frac{f_{l, m}^{\ast}}{A_{l, m}^{\ast}}. 
\end{eqnarray}


\section{Generalization of the pathway network to $N$ monomer network}
Our aim was to understand the behavior of the investigated network.

\begin{figure}
\begin{center}
\includegraphics[width=12cm]{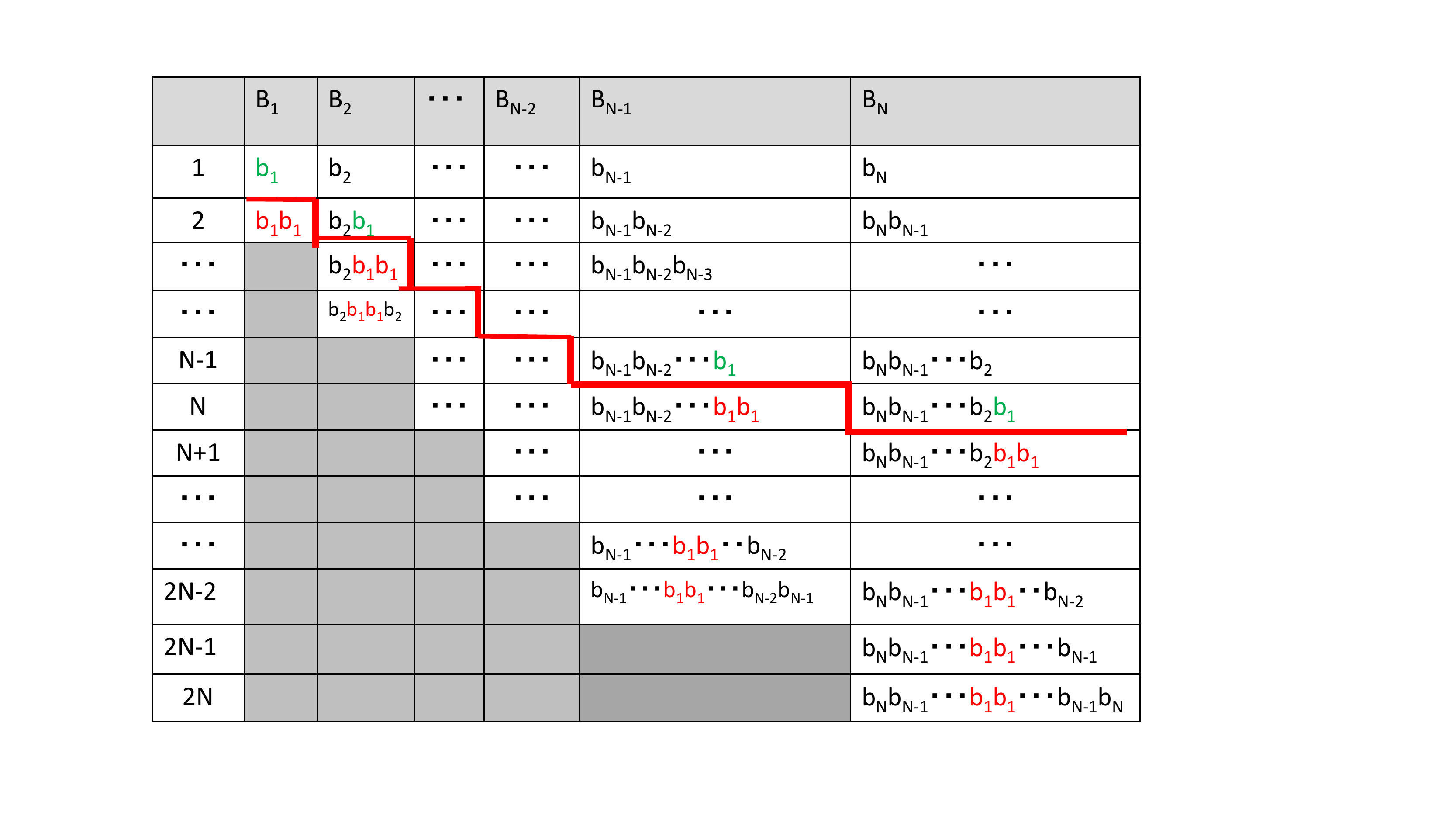}
\end{center}
  \caption{Diagram molecules belonging to the N monomer network.
      Rows represent the number of monomers in the complex. The column is the biggest monomer index number. The molecules are expressed with the complex formation consisting N monomers, like $(b_m-\cdots-b_1-b_1-\cdots-b_{l-m})$. }
\label{MoleculeOfN}
 \end{figure}

\begin{figure}
  \begin{center}
\includegraphics[width=12.5cm]{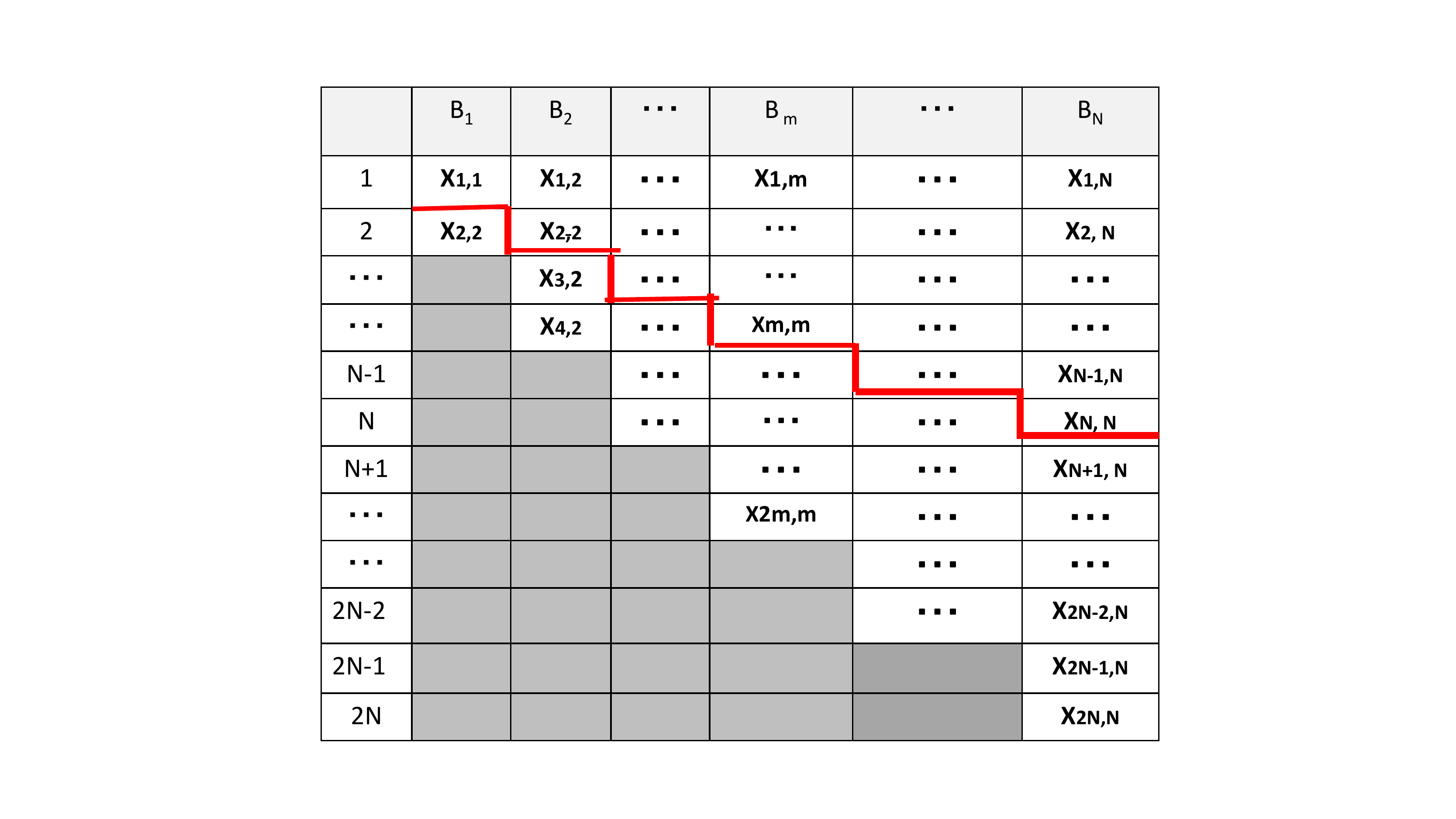}
\end{center}
  \caption{ Diagram of molecules in the N monomer network with the $X_{l,m}$ expression.
      The molecule $X_{l,m}$ has $l$ monomers and the biggest monomer index number is $m$. When $m\ge l$, the formation of  molecular $X_{l,m}$ is $(b_m-\cdots-b_{m-l+1})$. When $m < l$, the formation of molecular $X_{l,m}$ is $(b_m-\cdots-b_1-b_1-\cdots-b_{l-m})$. }
\label{NmodelFig}
 \end{figure}

The $N$ monomer network has $N(N+1)$ types of molecules. 
(Fig.\ref{MoleculeOfN}-Fig.\ref{NmodelFig}). The molecular concentration is expressed as $X_{l, m}(t)$, where $m$ is the biggest index number of a monomer in the molecule, while $l$ is the length of the molecule. 

%
\subsection{Classification of complexes to reaction groups}
\begin{figure}[h!]
\begin{center}
 \includegraphics[width=12cm]{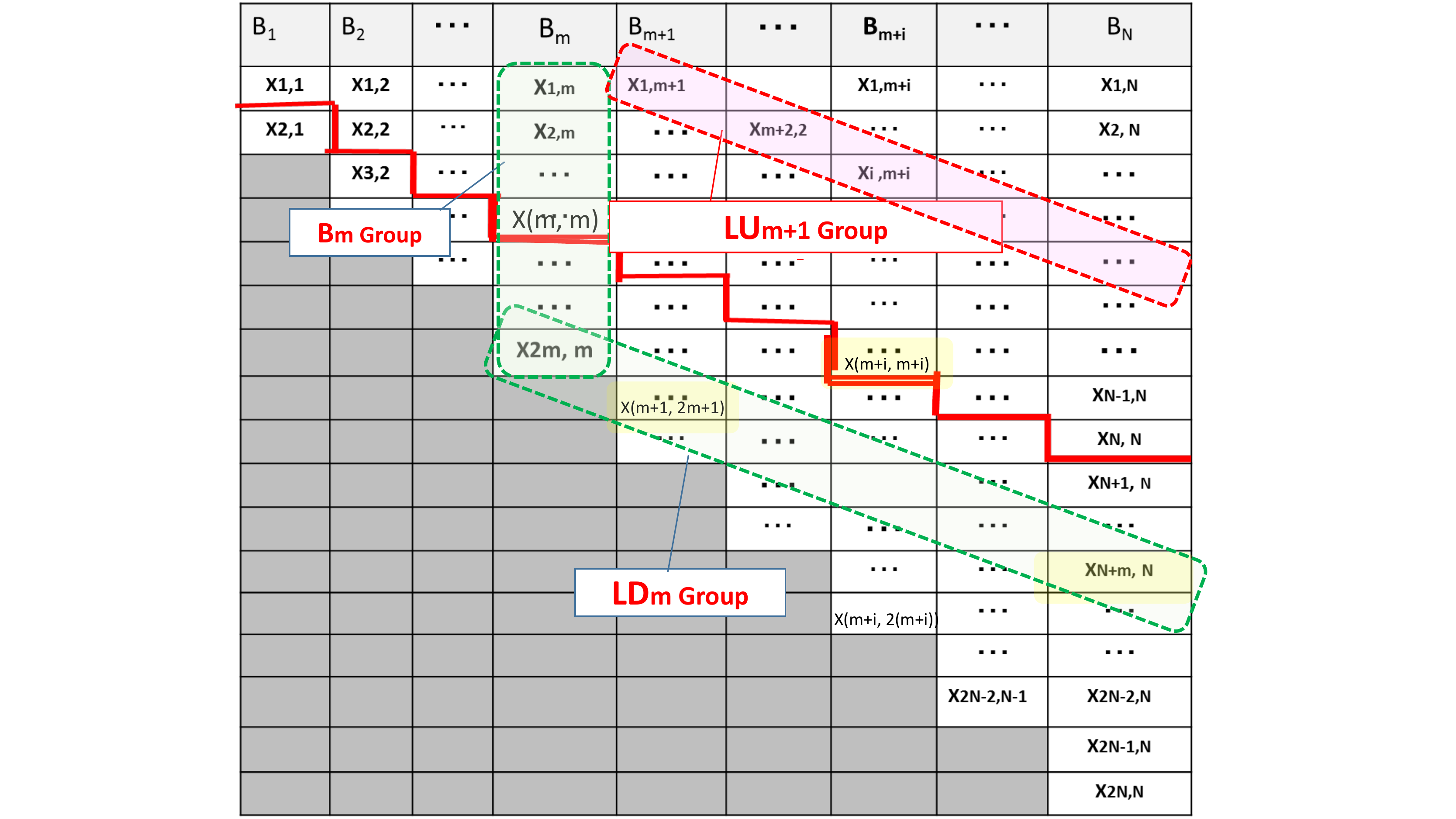}
 \end{center}
  \caption{Reaction groups. 
      The molecules in the ($b_N-\cdots-b_1-b_1-\cdots-b_N$) type network are classified into several reaction groups.}
\label{fig:4}
\end{figure}

We classified the molecules to several reaction groups according to the resulting reactions of the molecule $X_{l, m}$. (Fig.\ref{fig:4})

\begin{figure}[h]
\begin{center}
\includegraphics[width=12cm]{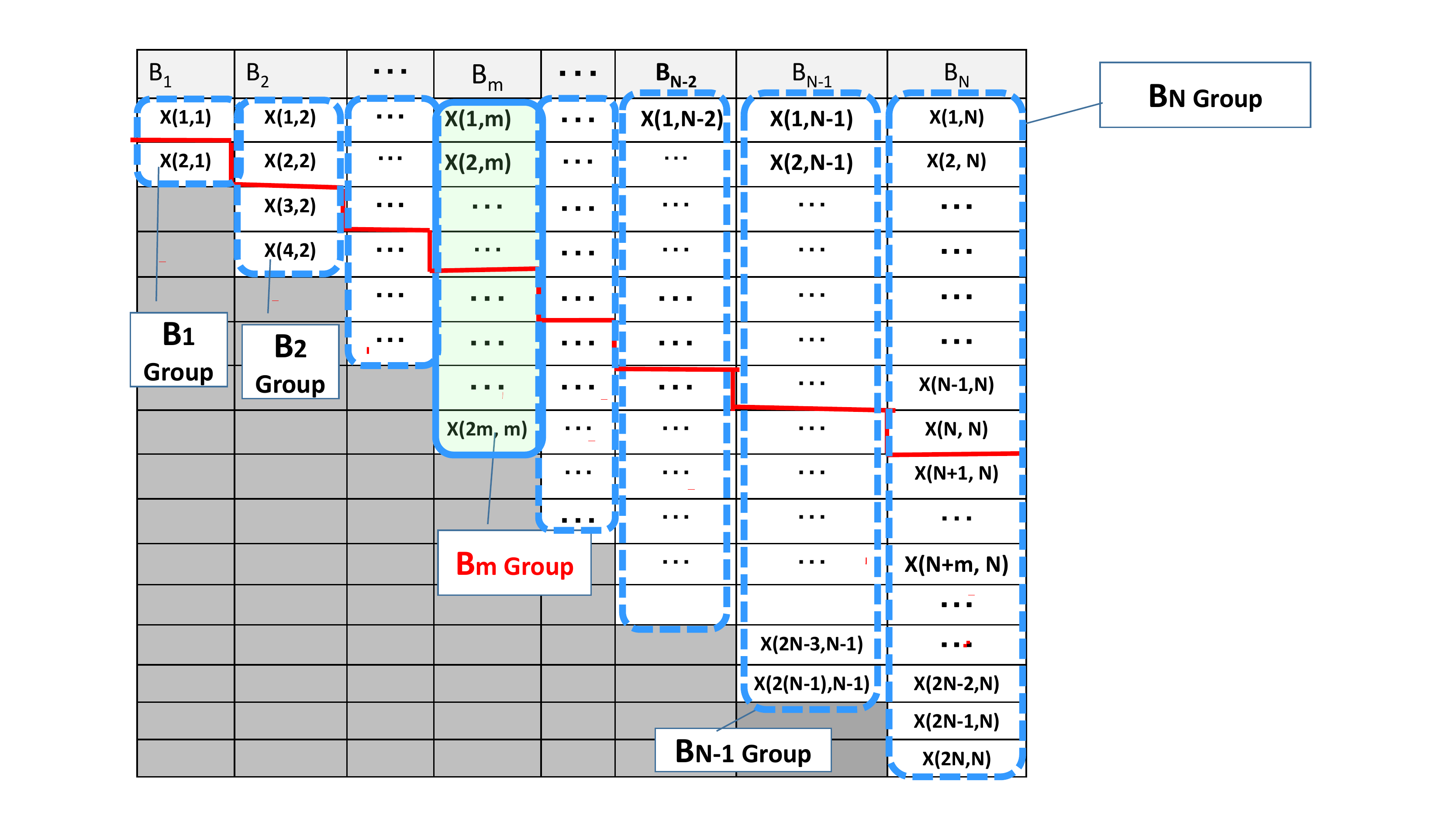}
\end{center}
  \caption{B groups. 
     The molecules in the ($b_N-\cdots-b_1-b_1-\cdots-b_N$) type network are classified to $N$ B groups according to the biggest monomer index number of the complex. For example, $B_m$ group has 2$m$ molecules, which have ($b_m-\cdots-b_i$) forms (monomer index $i$ is smaller or equal to $m$).}
\label{fig:5}
       \end{figure}

First, we classified the molecules to the N groups according to the biggest monomer index number in the molecule. The monomer with the biggest index number is the furthest from the central monomer $b_1$. We defined a reaction group as $B_i (i= 1, ..., N)$. Group $B_i $ consists of molecules, $X_{1,  i}: (b_i) $, $X_{2,i}: (b_i - b_{i-1})$, ...,  $X_{i, i }: (b_i -\cdots-b_{1})$,  $X_{i+1, i }: (b_i- \cdots-b_{1}- b_{1})$, ... , $X_{2i, i }: (b_i- \cdots-b_{1}- b_{1}-\cdots-b_i)$(Fig.\ref{fig:5}), the molecules in the group $B_i$ have the edge monomers $b_i (i= 1,\cdots, N)$, and $i$ is the biggest monomer index number among  the monomers included in the $B_i$ group molecule. 


In contrast to this, we classified network complexes into group $LU_j$ or group $LD_j$ according to the edge monomer index number, which is not the biggest index number in the molecule ($j = 1, ..., N$) .

\begin{figure}[h]
\begin{center}
 \includegraphics[width=12cm]{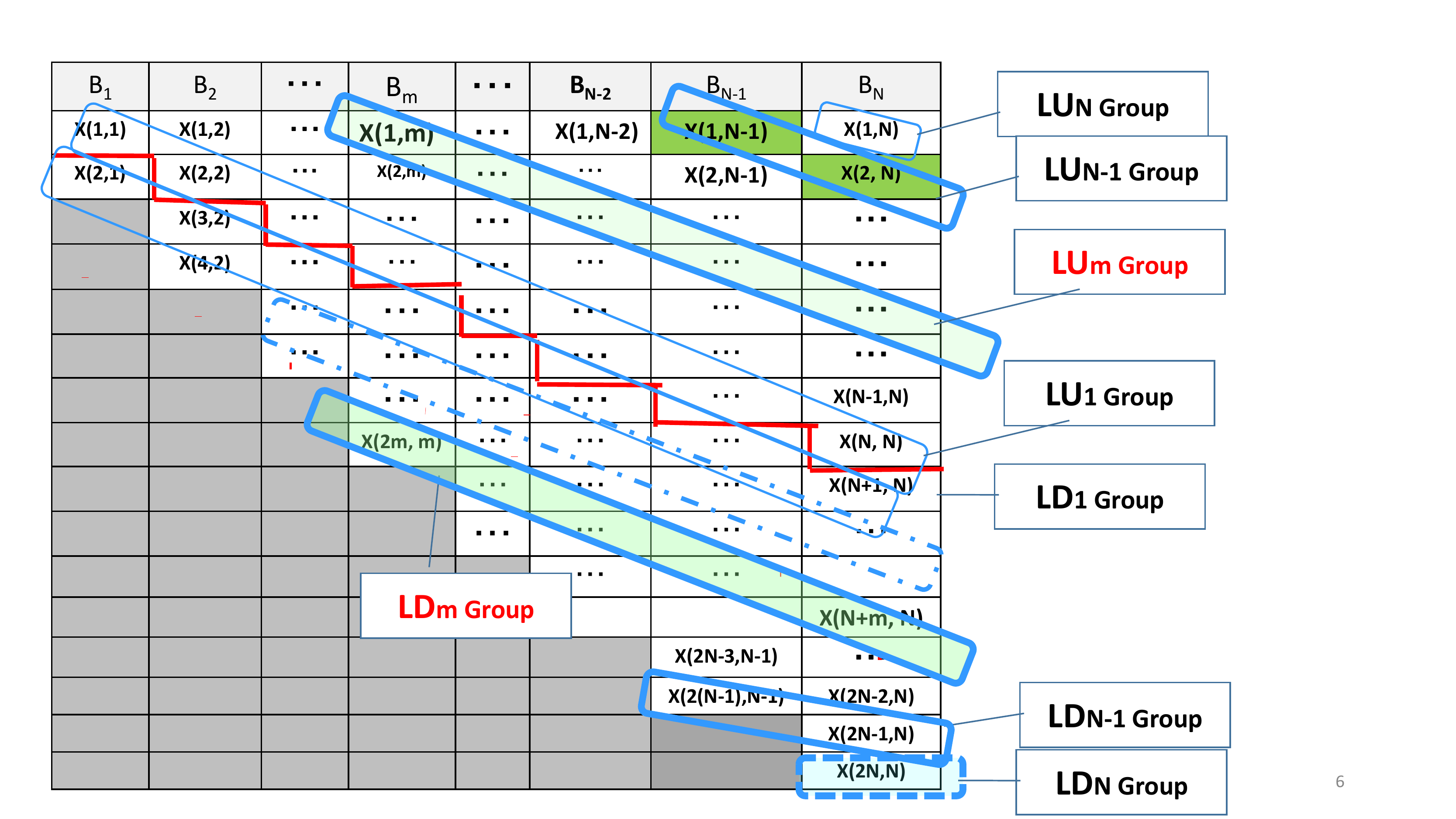}
 \end{center}
  \caption{
      The molecules in the ($b_N-\cdots-b_1-b_1-\cdots-b_N$) type network are classified to $N$ LU groups and $N$ LD groups, according to the edge monomer index number in the molecule, which is not the biggest monomer index number in the molecule. $LU_m$ group has $m$ molecules, which have $b_m$ monomer on the edge and no $b_1-b_1$ dimer. Their molecular form is ($b_i-\cdots-b_m$). The monomer index number $i$ is bigger than or equal to $m$. $LD_m$ group has $m$ molecules, which have $b_m$ monomer on the edge and $b_1-b_1$ dimer form. The molecular form is ($b_i-\cdots-b_1-b_1-\cdots-b_m$) (monomer index $i$ is smaller or equal to $m$).}
\label{fig:6}
 \end{figure}

Group $LU_j $ consists of molecules: $X_{1, j }: (b_j)$,  $X_{2, j+1}: (b_j+1- b_{j})$, ..., $X_{N-j+1, N }: (b_N -\cdots- b_{j})$ (Fig.\ref{fig:6}).
The molecules in the $LU_j$ group have monomer $b_j$ on the edge and no $b_1-b_1$ polymer.

Group $LD_j $ consists of molecules: $X_{2j, j}: (b_j \cdots b_1-b_1 \cdots b_j )$, $X_{2j+1, j+1}: (b_{j+1}b_j \cdots b_1-b_1 \cdots b_j )$, ..., $X_{N+j, N}: (b_N \cdots b_{j}\cdots b_1-b_1 \cdots b_j )$.
The molecules in the $LD_j$ group have monomer $b_j$ on the edge and $b_1-b_1$ polymer. 
%
 

\subsection{The law of mass conservation for the reaction groups}
\begin{figure}[h!]
  \begin{center}
 \includegraphics[width=12cm]{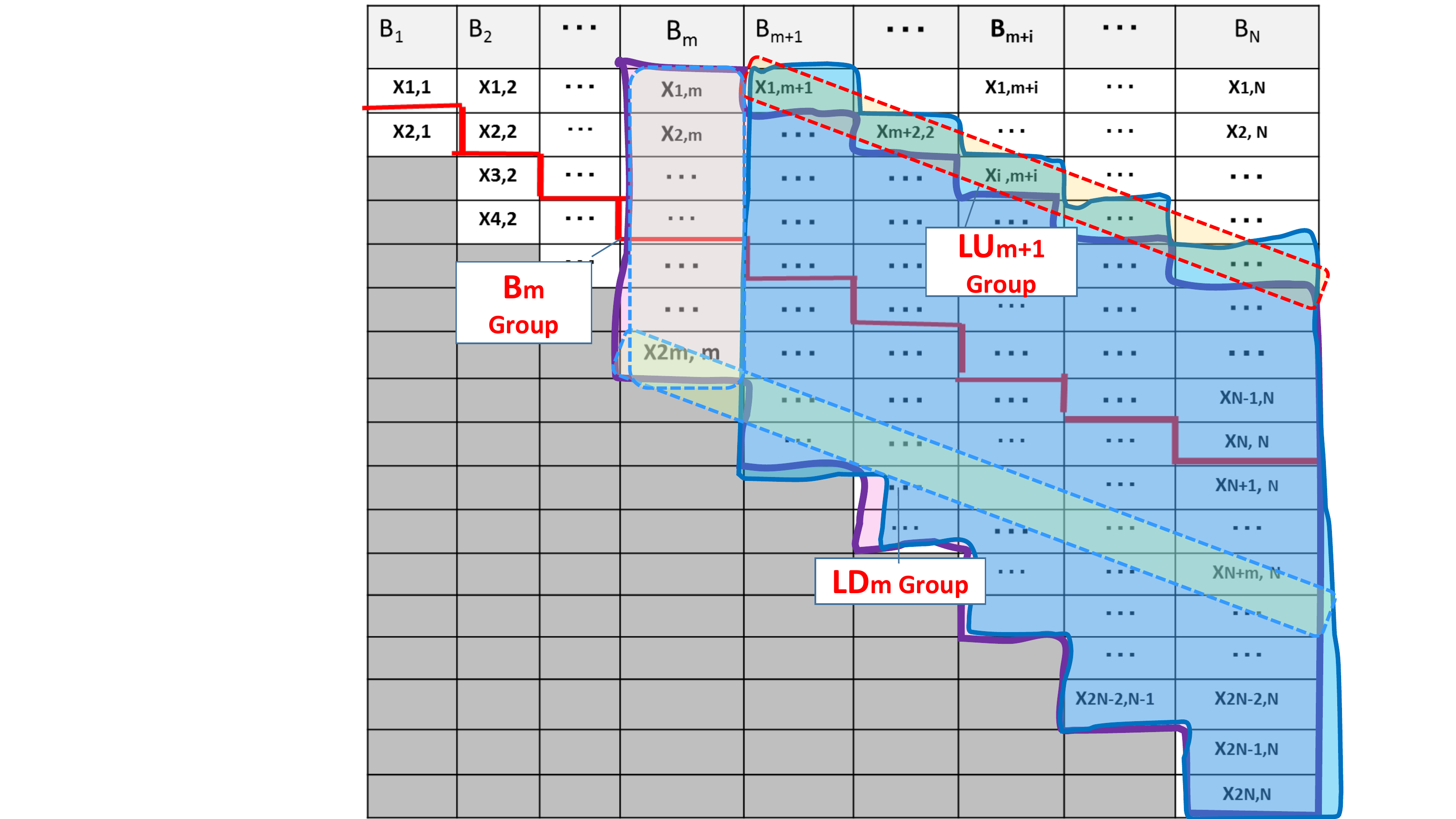}
 \end{center}
  \caption{
  N equations define the relationship between the $LU_{m+1}$ and $(B_m+LD_m)$ reaction groups, according to the law of mass conservation.
The molecules in the blue region contain $b_{m+1}$ monomers. The molecules in the purple region contain $b_m$ monomers.}
\label{MCLFig}
       \end{figure}

According to the law of mass conservation, we determined N equations for relationships between the $LU_{m+1}$ and $(B_m+LD_m)$ reaction groups. 
First, we found a total concentration of monomer $b_N$ in the network, which is equal to initial concentration of $[b_N(0)]$. 
$[b_N(0)]$ is derived as follows:
\begin{eqnarray}
\label{EqMCLn}
[b_N(0)]=\Sigma _{i=1}^{2N} X_{i, N } (t)+ X_{2N, N}(t).\hspace{1 cm}
\end{eqnarray} 
In the equation (\ref{EqMCLn}), $X_{2N,N} (t)$ is doubled, because the complex, X(2N,N) has two $b_N$ monomers. $[b_N (0)]$ is can also be written as follows:
\begin{eqnarray}
[b_N (0)] = \Sigma _{total}B_N + \Sigma _{total} LD_N.
\end{eqnarray} 
$\Sigma _{total}B_N$ is a total concentration of complexes in group $B_N$, while 
$\Sigma _{total}LD_N$ is the total concentration of complexes in group $LD_N$ 
(Fig. $\ref{MCLFig}$).

Following this, we derived the concentration of monomer $[b_{N-1}(0) ]$. In this network, the molecules that have $b_{N-1}$  monomers, are included in the $B_N$ group and $B_{N-1}$ group.
Each molecule in $B_{N-1}$ group contains $b_{N-1}$ monomers. Each molecule in $LD_{N-1}$ group contains two $b_{N-1}$ monomers.   No molecules in $LU_{N}$ group have any $b_{N-1}$ monomers. Therefore, we obtained the concentration of monomer $[b_{N-1} (0)]$ by the addition of $\Sigma _{total}B_{N-1}$, $\Sigma _{total}LD_{N-1}$, and $-\Sigma _{total}LU_N$ to the concentration of monomer $[b_{N} (0)]$, as follows:
\begin{eqnarray}
[b_{N-1} (0)]-[b_N(0)]=\Sigma _{total}B_{N-1}  + \Sigma _{total}LD_{N-1} -\Sigma _{total}LU_N.\hspace{1 cm}
\end{eqnarray} 
If the initial concentration $[b_{m+1}(0)]$ of $b_{m+1}$ molecule is already known, we can calculate the initial concentration $[b_{m}(0)]$ of $b_{m}$ molecule as follows:
\begin{eqnarray}
\label{EqMCLnh}
[b_{m} (0)]-[b_{m+1}(0)]=\Sigma _{total}B_{m}  + \Sigma _{total}LD_{m} -\Sigma _{total}LU_{m+1}. \hspace{1 cm}
\end{eqnarray} 
Therefore, we were able to obtain $N$ mass conservation relationships between the reaction groups.
We determine the sum of the equation ($\ref{EqMCLnh}$), from $m$ to $N$. Left-side terms of this equation cancel each other, leaving only $[b_m(0)]-[b_N (0)]$:
\begin{eqnarray}
\label{EqMCLh}
[b_{m} (0)]-[b_{N}(0)]&=&\Sigma _{j=m}^{N} ( [b_{j} (0)]-[b_{j+1} (0)] ), \nonumber
\\ &=&\Sigma _{j=m}^{N-1} (\Sigma _{total}B_{j}  + \Sigma _{total}LD_{j} -\Sigma _{total}LU_{j+1}). \hspace{1 cm}\nonumber 
\end{eqnarray} 
Followed with:
\begin{eqnarray}
\label{EqMCLh1}
[b_{m} (0)] &=&\Sigma _{j=m}^{N} ( \Sigma _{total}B_{j}  + \Sigma _{total}LD_{j} )  -\Sigma _{j=m}^{N-1}\Sigma _{total}LU_{j+1}.   \hspace{1 cm}
\end{eqnarray}

\subsection{The law of mass action for the reaction groups}
As previously described, mass preservation relationships between the reaction groups, $B$, $LU$, and $LD$ groups are present. Here, we introduce the law of mass action for the reaction groups and their group concentrations at previous work. We show how to derive  the mass action laws and their solutions in the appendix A.

%
We define the parameter $\xi_ {m}(t)$ as the total concentration of the molecules belonging to the $LU_{m}$ group.
The parameter $\eta_{m}(t)$ is defined as the total concentration of molecules in $B_{m}$ and $LD_{m}$ groups.

\begin{eqnarray}
\left\{
\begin{array}{l}
\xi_ {m}(t)= \Sigma _{total}LU_{m}(t), \\
\eta_{m}(t)=\Sigma _{total}(B_m(t)+LD_m(t)).
\end{array}
\right.
\end{eqnarray}

Afterward, equations ($\ref{EqMCLnh}$)-($\ref{EqMCLh1}$) are rewritten using $\xi_{m+1}(t)$ and $\eta_m(t)$:
\begin{eqnarray}
\label{EqMCLm}
[b_{m}(0)]-[b_{m+1}(0)]= \eta_m(t) -\xi_{m+1}(t).  
\end{eqnarray}

The LU reaction group concentrations are summarized as follows:
\begin{eqnarray}
\xi_{N}(t)&=\left\{ \begin{array}{ll} 
\frac{\xi_{{N} +}^{\ast}- C_{N}\xi_{{n} -}^{\ast}e^{-\beta_{n}t}}{1-C_{n}e^{-\beta_{n}t}} &(l_{N}>0, m=N),\\
\frac{[b_{N}(0)]}{k_{N}[b_{N}(0)]t+1},& (l_{N}=0$ and $[b_{N-1}(0)]=[b_{N}(0)], m=N).\\
\end{array} \right.
\end{eqnarray}

\begin{eqnarray}
\xi_{m}(t)&=\left\{ \begin{array}{ll} 
\frac{\xi_{{m} +}^{\ast}- C_{m}\xi_{{m} -}^{\ast}e^{-\beta_{m}t}}{1-C_{m}e^{-\beta_{m}t}}& (l_{m}>0, m:1\downarrow N-1),\\
\frac{[b_{m}(0)]}{k_{m}[b_{m}(0)]t+1}& (l_{m}=0$ and $[b_{m-1}(0)]=[b_{m}(0)],m:1\downarrow N-1).\\
\end{array} \right.
\end{eqnarray}

\begin{eqnarray}
\xi_{1}(t)&=\left\{ \begin{array}{ll} 
\frac{\xi_{1 +}^{\ast}- C_{1}\xi_{1 -}^{\ast}e^{-\beta_{1}t}}{1-C_1e^{-\beta_{1}t}}&(l_{1}>0, m=1),\\
\frac{[b_1(0)]}{k_1[b_1(0)]t+1}& (l_{1}=0, m=1).\\
\end{array} \right.
\end{eqnarray}
LU group concentrations, $\xi_{N},...,\xi_{m},...,\xi_{1}$, are independent of each other and have strict solutions, which converge to equilibrium solutions when time t tends to $+\infty$.\\
The total concentration of $B_m$ and $LD_m$ groups $\eta_m$ is derived from the equation ($\ref{EqMCLnh}$), as follows:
\begin{eqnarray}
\left\{\begin{array}{ll} 
\eta_{N}(t)&=[b_N(0)], \hspace{1cm}(m=N),\\
\eta_{m}(t)&=[b_{m}(0)]-[b_{m+1}(0)]+\xi_{m+1}(t), \hspace{1cm}(m=1,\cdots, N-1).
\end{array} \right.
\end{eqnarray}
Here, we showed that the total concentration of $LU_m$ group, or total concentrations of $LD_m$ and $B_m$ groups, can be solved. They converge to the equilibrium solutions when time t tends to $+\infty$.

We regarded the reaction group concentrations as a priori upper boundaries of the concentration of each molecular in the group.
Afterward, we demonstrated that each molecule in a group converges to a stable solution.

%
\section{Concentrations of complexes in the network }
All evolution equations of the complexes in the pathway network were derived.
We show that  ODEs of all concentrations of the molecules $X_{l,m}(t) $ in the network have the following shape:
 \begin{eqnarray}
\label{EqCmpX}
\frac{d X_{l,m}(t) }{dt} = -A_{l,m}(t)X_{l,m}(t)+f_{l,m}(t).  \hspace{1 cm}
\end{eqnarray}
First, we solved the homogeneous differential equation, $\frac{d X_{l,m}(t) }{dt} = -A_{l,m}(t)X_{l,m}(t)$, and obtained the solution $X_{l,m}(t) = X_{l,m}(0) \  \mathrm{e}^{-\int_0^tA_{l,m}(s)ds}$. 
These equations were solved by varying the parameters.
\begin{eqnarray}
\label{SolCmpX}
 X_{l,m}(t) =  \mathrm{e}^{-\int_0^tA_{l,m}(s)ds}+\int_0^t  \mathrm{e}^{-\int_0^sA_{l,m}(u)du} f_{l,m}(s)ds. \hspace{1 cm}
\end{eqnarray}
Afterward, all complex concentrations were shown to be integrable.

We set equilibrium values for molecules $X_{l,m}$, $A_m(t)$, and $f_m(t)$ as $X^{\ast}_{l,m}$ , $A_{l,m}^{\ast}$, and $f_{l,m}^{\ast}$.
We defined the parameter $Y(t) = X_{l,m} (t)-X^{\ast}_{l,m}$:
 \begin{eqnarray}
\frac{d Y_{l,m}(t) }{dt} = -A_{l,m}(t)Y_{l,m}(t)+\left(f_{l,m}(t)-\frac{f^{\ast}_{l,m}}{A_{l,m}^{\ast}} A_{l,m}(t)\right). \hspace{1 cm}
\end{eqnarray}
We defined $g_{l,m}(t)=f_{l,m}(t)-\frac{f^{\ast}_{l,m}}{A_{l,m}^{\ast}} A_{l,m}(t)$ and $a_{l,m}(t)= \int_T^{t}A_{l,m}(s)ds$.

\begin{eqnarray}
 Y_{l,m}(t) &=&  \mathrm{e}^{-a(t)+a(T)}Y_{l,m}(T)+\int_T^t  \mathrm{e}^{a(s)-a(t)} g_{l,m}(s)ds \hspace{1 cm}\nonumber\\
		&\le& \mathrm{e}^{a(T)}|Y_{l,m}(T)|\mathrm{e}^{-A_{l,m}^{\ast}(t-T)}+\int_T^t \mathrm{e}^{-A_{l,m}^{\ast}(s-T)}\mathrm{e}^{-at}\mathrm{e}^{-\delta s}ds,
\end{eqnarray}
where $a(t) \ge A_{l,m}^{\ast}(t-T)$.\\
Because $ A_{l,m}(t) = k_i \xi_i(t) +k_j \xi _j(t) + l_i +l_j$ and integral of $\xi_i(t)$ is written as follows:
\begin{eqnarray}
\int_T^t \xi_m(s)ds &= &\left[\xi_{m+}^{\ast}s+\frac{1}{B_m} \log\left(1-C_m\right) \mathrm{e}^{-B_m s}\right] _T^t \hspace{1 cm} \nonumber\\
&=&\xi_{m+}^{\ast}(t-T)+\frac{1}{B_m} \log \left(\frac{1-C_m \mathrm{e}^{-B_mt}}{1-C_m \mathrm{e}^{-B_mT}}\right),
\end{eqnarray}
where $B_m=k_m (\xi_{m+}^{\ast}-\xi_{m-}^{\ast})$, $C_m=\frac{[b_m(0)]-\xi_{m+}^{\ast}}{[b_m(0)]-\xi_{m-}^{\ast}}$.
When time $t$ tends to $\infty$, $Y_{l,m}(t)$ tends to $0$:
\begin{eqnarray}
\lim_{t \to \infty} X_{l,m}(t) = \frac{f_{l,m}^{\ast}}{A_{l,m}^{\ast}}.
\end{eqnarray}

\subsection{The concentration and stability of complexes in LU groups }
%
We calculated the concentrations of the molecules by starting with the complex $X_{1, 2} \subset  LD_1, B_1$
followed by solving ODEs for complexes $ X_{3, 2} \subset B_2, X_{4, 3} \subset B_3,\cdots., X_{N-m+1, N} \subset B_{N} $  in  $LD_1 $ group, step by step. Afterward, we solved the ODEs of complexes in $LD_2$, $\cdots$, $LD_N$ group.

Additionally, we began with the molecules in the $LU_N$ group, and solved the concentrations of molecules from $LU_{N-1}, \cdots, LU_1$, $LD_1,\cdots, LD_N$ groups, step by step.  

\subsection{Complex concentrations in $LU_{N}$ group}
First, we started with $LU_N$ group and calculated the complex concentrations of $LU_{N-1}, \cdots$, $LU_{2}$, $LU_{1}$. 
$LU_{N}$ group has only one complex, $X_{1,N}$, and its concentration $X_{1, N}(t)$ is equal to $LU_N$ group solution, $\xi_N(t)$.
\begin{eqnarray}
\label{EqXLUN}
 X_{1, N}(t) &=& \xi_{N}(t).  
\end{eqnarray}
Its equilibrium is obtained.
\begin{eqnarray}
\label{EqEqXLUN}
 X_{1, N}^{\ast} &=& \xi_{N}^{\ast}.  
\end{eqnarray}
\subsection{Concentrations of molecules in $LU_{N-1}$($N \geq 2$) group}
We considered the concentrations of molecules in $LU_{N-1}$($N \geq 2$) group, $X_{i, N+i-2}(t) \subset B_{i} $(i = 1, 2 ). 
\begin{eqnarray}
\label{EqXLUN1}
\frac{d X_{i, N+i-2}(t) }{dt} = -A_{i, N+i-2} (t)X_{i, N+i-2}(t)+f_{i, N+i-2}(t) \hspace{ 0.3 cm}(i=1,  2)
\end{eqnarray}
where
\begin{eqnarray}
A_{i, N+i-2}(t)=
k_{N}\xi_{N}(t)+k_{N-1}\eta_{N-2}(t)+l_{N}+l_{N-1} \hspace{0.3 cm}(i =1,2).\nonumber
\end{eqnarray}
\begin{eqnarray}
f_{i, N+i-2}(t)=\left\{ \begin{array}{ll} 
&l_{N}\xi_{N-1}(t)+l_{N-1}\eta_{N-1}(t) \hspace{0.3cm}(i =1),\\\\
& k_{N}\xi_N(t)\xi_{N-1}(t) \hspace{ 0cm}
+l_{N-1} (\eta_N(t)- \xi_N(t) ) 
\hspace{0.3cm}(i = 2).
\end{array}\right.
\end{eqnarray}

Its equilibrium solution is
\begin{eqnarray}
\label{SolXLUN1}
\lim_{t \to \infty}X_{i, m+i-1}^{\ast} = \frac {f^{\ast}_{i, m+i-1}} {A^{\ast}_{i, m+i-1} } \hspace{ 0.4 cm}(i=1, 2),
\end{eqnarray}
where
\begin{eqnarray}
A_{i, N+i-2}^{\ast}=
k_{N}\xi_{N}^{\ast}+k_{N-1}\eta_{N-2}^{\ast}+l_{N}+l_{N-1} \hspace{0.4cm}(i =1,2), \nonumber
\end{eqnarray}
\begin{eqnarray}
f_{i, N+i-2}^{\ast}=\left\{ \begin{array}{ll} 
&l_{N}\xi_{N-1}^{\ast}+l_{N-1}\eta_{N-1}^{\ast} \hspace{0.3cm}(i =1),\\\\
& k_{N}\xi_N^{\ast}\xi_{N-1}^{\ast} +l_{N-1} ( \eta_N^{\ast}- \xi_N^{\ast} ) 
\hspace{0.3 cm}(i = 2).
\end{array}\right.
\end{eqnarray}
\subsection{Mmolecule concentraion in the $LU_{m}$ group $(m = 2, \cdots, N-2)
$}
Afterward, we considered the concentrations of molecules in $LU_m$ group, specifically, of the $LU_{m}$ group complex, $X_{i, m+i-1}(t) \subset B_{i} (i = 1, \cdots, N-m+1)$. Starting from the mass conservation laws and mass action laws, we obtained the reaction ODEs for $X_{i, m+i-1}$ molecule as follows:
\begin{eqnarray}
\label{EqXLUm1}
\frac{d X_{i, m+i-1}(t) }{dt} = -A_{i, m+i-1} (t)X_{i, m+i-1}(t)+f_{i, m+i-1}(t) \nonumber\\
\hspace{ 0.7 cm}(i=1, \cdots, N-m+1)
\end{eqnarray}
where
\begin{eqnarray}
&A_{i, m+i-1}(t)=\left\{ \begin{array}{ll} 
&k_{m+1}\xi_{m+1}(t)+k_{m}\eta_{m-1}(t)+l_{m+1}+l_m \hspace{0.3cm}(i =1),\nonumber\\ \\  
&k_{m+i}\xi_{m+i}(t)+k_{m}\eta_{m-1}(t)+\sum_{j=0}^{i}  l_{m+j} 
\hspace{ 0.4cm}(i = 2, \cdots, N-m), \\\\
&k_{m}\eta_{m-1}(t) + \sum_{j=0}^{N-m} l_{m+j}  \hspace{0.3cm}(i = N-m+1),
\end{array} \right.
\end{eqnarray}
\begin{eqnarray}
f_{i, m+i-1}(t)=\left\{ \begin{array}{ll} 
&l_{m+1}\xi_{m}(t)+l_{m}\eta_{m}(t) \hspace{0.3cm}(i =1),\\\\
%
&l_{m+i}(\xi_m(t)-\sum_{j=1}^{i-1}X_{j,m+j-1}(t))
+l_{m}(\eta_{m+i}(t)- \sum_{j= 1}^{i-1}X_{j,m+i-1}(t)) 
\nonumber\\&+\sum_{j= 1}^{i-1}k_{m+j}X_{i-j, m+i-1}(t)X_{j,m+j-1}(t)
\hspace{0.5cm}(i = 2, \cdots, N-m),\\\\
&\sum_{j = 1}^{N-m} k_{m+j}X_{j,m+j-1}(t)X_{ N-m+1-j,N }(t) +l_{m} ( \eta_N(t)-\sum_{j=1}^{N-m} X _{j, N}(t) ) \hspace{ 0cm}\nonumber\\
&
\hspace{1cm}(i = N-m+1).\\
\end{array}\right.
\end{eqnarray}

Its equilibrium solution is
\begin{eqnarray}
\label{SolXLUm1}
\lim_{t \to \infty}X_{i, m+i-1}(t) = \frac {f^{\ast}_{i, m+i-1}} {A^{\ast}_{i, m+i-1} } 
\hspace{ 1 cm}(i=1, \cdots, N-m+1), 
\end{eqnarray}
where
\begin{eqnarray}
&A_{i, m+i-1}^{\ast}=\left\{ \begin{array}{ll} 
&k_{m+1}\xi_{m+1}^{\ast}+k_{m}\eta_{m}^{\ast}+l_{m+1}+l_{m} \hspace{0.3 cm}(i =1),\nonumber\\ \\
&k_{m+i}\xi_{m+i}^{\ast}+k_{m}\eta_{m-1}^{\ast} +\sum_{j=m }^{m+i}l_j 
\hspace{ 0.3cm}(i = 2, \cdots, N-m),\\\\
&k_{m}\eta_{m-1}^{\ast} +\sum_{j= 0}^{N-m}l_{m+j} \hspace{0.3cm}(i = N-m+1),
\end{array} \right.
\end{eqnarray}
\begin{eqnarray}
f_{i, m+i-1}^{\ast}=\left\{ \begin{array}{ll} 
&l_{m+1}\xi_{m}^{\ast}+l_{m}\eta_{m}^{\ast}  \hspace{0.3cm}(i =1),\nonumber\\\\
&  
l_{m}(\eta_{m+i}^{\ast}- \sum_{j= 1}^{i-1}X_{j,m+i-1}^{\ast})+l_{m+i}(\xi_{m}^{\ast}- \sum_{j= 1}^{i-1}X_{j, m+j-1}^{\ast})\\
&+ \sum_{j= 1}^{i-1}k_{m+j}X^{\ast}_{i-j, m+i-1}X^{\ast}_{j, m+j-1} 
\hspace{0.3cm}(i = 2, \cdots, N-m-1),\\\\
 &\sum_{j = 1}^{N-m} k_{m+j}X^{\ast}_{j,m+j-1}X^{\ast}_{N-m-j+1,N}
+l_{m} ( \eta_N ^{\ast}-\sum_{j=1}^{N-m} X ^{\ast}_{j,N})
\\ & \hspace{4cm}(i = N-m+1).\\
\end{array}\right.
\end{eqnarray}
\subsection{Concentration of $LU_{1}$ group complexes}
Reaction ODE of $LU_1$ complexes, $X_{i, i} \subset B_i (i = 1, \cdots, N)$, is written as follows:

\begin{eqnarray}
\label{EqX11}
\frac{d X_{1,1}(t) }{dt} = -A_{1, 1}(t)X_{1,1}(t)+f_{1,1 }(t),
\end{eqnarray}
where
\begin{eqnarray} 
\left\{ \begin{array}{ll} 
A_{1, 1}(t)=2k_{1}\xi_{1}(t)+2k_{2}\xi_{2}(t)+\sum_{j = 1}^{2} l_{j},\\ 
f_{1, 1}(t)= l_1 \eta_1(t)+l_2 \xi_1(t) .\nonumber
\end{array} \right.
\end{eqnarray}
Therefore, the equilibrium solution of $X_{i,i }(t) $ is
\begin{eqnarray}
\label{EquiXn}
X^{\ast}_{1, 1} = \lim_{t \to \infty}X_{1, 1}(t) 
=\frac{f_{1, 1}^{\ast}}{A_{1, 1}^{\ast}} ,
\end{eqnarray}
where
\begin{eqnarray}
\left\{ \begin{array}{ll} 
A_{1, 1}^{\ast} =2k_{1}\xi_{1}^{\ast}+2k_{2}\xi_{2}^{\ast}+\sum_{j = 1}^{2} l_{j},\\ 
f_{1, 1}^{\ast}= l_1 \eta_1^{\ast}+l_2 \xi_1^{\ast} .
\end{array} \right.
\end{eqnarray}

When i > 1,
\begin{eqnarray}
\label{EqXii}
\frac{d X_{i,i}(t) }{dt} = -A_{i, i}(t)X_{i,i}(t)+f_{i,i }(t) , (i = 2, \cdots,  N-1) ,
\end{eqnarray}
where
\begin{eqnarray} 
\left\{ \begin{array}{ll} 
A_{i, i}(t)=2k_{1}\xi_{1}(t)+k_{i+1}\xi_{i+1}(t)+\sum_{j = 1}^{i + 1} l_{j}, \\ 
f_{i, i}(t)=\sum_{j = 1}^{i-1} k_{j+1} X_{j, j}(t) X_{i-j, i} (t)+ l_{i+1} \{\xi_1(t)-\sum_{j = 1}^{i-1}X_{j, j}(t)\} 
\\ \hspace{2cm} + l_1\{ [b_i(0)] - [b_{i+1}(0)] + \xi_{i+1}(t)-\sum_{j=1}^{i-1}X_{j, i}(t)\}. \nonumber
\end{array} \right.
\end{eqnarray}
Therefore, the equilibrium solution of $X_{i,i }(t) $ is
\begin{eqnarray}
\label{EquiXn}
X^{\ast}_{i, i} = \lim_{t \to \infty}X_{i, i}(t) 
=\frac{f_{i, i}^{\ast}}{A_{i,  i}^{\ast}} ,  
\end{eqnarray}
where
\begin{eqnarray}
\left\{ \begin{array}{ll} 
A_{i, i}^{\ast} =2k_{1}\xi_{1}^{\ast}+k_{i+1}\xi_{i+1}^{\ast}+\sum_{j = 1}^{i + 1} l_{j}, \\
f_{i, i}^{\ast}=\sum_{j = 1}^{i-1} k_{j+1} X^{\ast}_{j, j} X^{\ast}_{i-j, i} + l_{i+1} \{\xi_1^{\ast}-\sum_{j = 1}^{i-1}X^{\ast}_{j, j}\} 
\\ \hspace{2cm} + l_1\{ [b_i(0)] - [b_{i+1}(0)] + \xi_{i+1}^{\ast}-\sum_{j=1}^{i-1}X^{\ast}_{j, i}\}.
\end{array} \right.
\end{eqnarray}

Finally, for the molecules belonging to the $LU_1$, $X_{N, N} \subset B_N $, reaction ODE of this molecule is written:
\begin{eqnarray}
\label{EqXnn}
\frac{d X_{N, N}(t) }{dt} = -A_{N, N}(t)X_{N, N} (t)+f_{N, N }(t) ,
\end{eqnarray}
where 
\begin{eqnarray} 
\left\{ \begin{array}{ll} 
A_{N, N}(t)=2k_{1}\xi_{1}(t)+\sum_{j = 1}^{n} l_{j},\\
f_{N, N}(t)=\sum_{j = 1}^{N-1} k_{j+1} X_{j, j}(t) X_{N-j, N} (t) + l_1( [f_N(0)] -\sum_{j=1}^{N-1}X_{j, N}(t)).\nonumber
\end{array} \right.
\end{eqnarray}
Therefore, the equilibrium solution of $X^{\ast}_{N, N } = \lim_{t \to \infty}X_{N, N}(t)$ is
\begin{eqnarray}
\label{EquiXn}
X^{\ast}_{N, N} =\frac{f_{N, N}^{\ast}}{A_{N, N}^{\ast}} ,
\end{eqnarray}
where
\begin{eqnarray}
\left\{ \begin{array}{ll} 
A_{N, N}^{\ast}=2k_{1}\xi_{1}^{\ast}+\sum_{j = 1}^{N} l_{j},\\ \\
f_{N,N}^{\ast}=\sum_{j = 1}^{N-1} k_{j+1} X_{j, j}(t) X^{\ast}_{N-j, N}+ l_1( [b_N(0)] -\sum_{j=1}^{N-1}X^{\ast}_{j, N}).\nonumber
\end{array} \right.
\end{eqnarray}

\subsection{Concentrations of complexes  in LD groups }
We demonstrated that the complex concentrations in LU groups are integrable and converge to the equilibrium. Following this, we solved the concentrations of LD group complexes. Here, we initially determined the concentrations of $X_{2, 1} \subset  LD_1, B_1$ complex, followed by ODEs for complexes $ X_{3, 2} \subset B_2, X_{4, 3} \subset B_3,\cdots,  X_{ N-m+1, N} \subset B_{N} $  in  $LD_1 $ group. Finally, we solved the ODEs of complexes belonging to $LD_2$ group, $\cdots$, $LD_N$ group, step by step.
%
\subsection{Concentration of complexes in $LD_{1}$ group}
$LD_{1}$ group contains $N$ complexes, $X_{2, 1}, X_{3, 2}, \cdots,X_{N+1, N}$. This is a group with special characteristics, as the complexes belonging to this group have $b_{1}-b_{1}$ connection on the edge. Any complex in $LD_1$ can react with every complex in $LU_2$ group. 

First, we considered the concentration of $LD_{1}$ group complex, $X_{ i+1, i}(t) \subset B_{i} (i = 1 , \cdots,  N)$. From the law of mass conservation and the law of mass action, we obtained reaction ODEs of molecule $X_{i+1, i}$ as follows:
\begin{eqnarray}
\label{EqXLD1}
\frac{d X_{i+1, i}(t) }{dt} = -A_{i+1, i}(t)X_{i+1, i}(t)+f_{i+1, i}(t) \hspace{ 0.3 cm}(i = 1 , \cdots,  N),
\end{eqnarray}
where 
\begin{eqnarray}
&A_{i+1, i}(t)=\left\{ \begin{array}{ll} 
&2k_{2}\xi_{2}(t)+2l_{2}+l_{1} \hspace{0.cm}(i =1), \nonumber\\\\
&k_{2}\xi_{2}(t)+k_{i+1}\xi_{i+1}(t)+l_1+2l_{2}+\sum_{j= 3}^{i+ 1}l_j  \hspace{0.3cm}
(i = 2 , \cdots,  N-1),\\\\
&k_{2}\xi_{2}(t)+l_{1}+2l_{2}+\sum_{j= 3}^{N}l_j \hspace{0.3cm}(i = N),\nonumber\\\\
\end{array} \right. 
\end{eqnarray}
\begin{eqnarray}
&f_{i+1, i}(t)=\left\{ \begin{array}{ll} 
&2k_{1}X_{1, 1}^{2}(t) +l_2 ( \eta_1(t) -X_{1, 1}(t)) \hspace{0.5cm}(i =1),\nonumber\\\\
&2k_{1}X_{1, 1}(t) X_{i, i}(t)+ k_2X_{2,1}(t)X_{i-1,i}(t)+\sum_{j= 1}^{i- 1}k_{j+1}X_{j+1, j}(t) X_{i-j, i}(t) \\
&+l_2\{\eta_{i}(t) -\sum_{j=1}^{i}X_{j, i}(t)\} 
 \hspace{ 0cm}+l_{i+1} ( \eta_1(t)-\sum_{j=1}^2 X_{j, 1}(t) -\sum_{j = 1}^{i-1}X_{j+1, j}(t)  )
\\&
\hspace{0.3cm}(i = 2 , \cdots,  N-1), \\
\\&2k_{1}X_{1, 1}(t) X_{N, N}(t)+ 2k_2X_{2,1}X_{i-1,i}+\sum_{j = 2}^{N-1} k_{j+1}X_{j+1, j}(t) X_{N-j, N}(t)
\\& +l_{2}  [b_N(0)]-\sum_{i=1}^{N} X_{i,N}(t)   \hspace{1cm} (i = N). \\
\end{array} \right.
%
\end{eqnarray}
Deformation of equation (\ref{EqXLD1}) is:
\begin{eqnarray}
\label{DefEqXiip1}
&\frac{d }{dt}  \{ X_{i+1, i}(t)- \frac{b_{i+1, i}^{\ast}} {A_{i+1, i}^{\ast}} \} &= -A_{i+1, i}(t) \{ X_{i+1, i}(t)- \frac{b_{i+1, i}^{\ast}} {A_{i+1, i}^{\ast}} \}+ f_{i+1, i}(t) 
-\frac{b_{i+1, i}^{\ast}} {A_{i+1, i}^{\ast} } A_{i+1, i}(t) \nonumber\\&&
\hspace{4cm}(i = 1 , \cdots,  N). 
\end{eqnarray}
Let $Y_{i+1, i}(t)$ be a difference between $X_{i+1, i}(t)$ and its equilibrium solution, $X^{\ast}_{i+1, i}= \frac{f_{i+1, i}^{\ast}} {A_{i+1, i}^{\ast}} $.
$Y_{i+1, i}(t)$ is equal to $X_{i+1, i}(t)- X^{\ast}_{i+1, i}$. \\
Then, 
\begin{eqnarray}
& \frac{d }{dt} Y_{i+1, i}(t) &= -A_{i+1, i}(t)Y_{i+1, i}(t)  + f_{i+1, i}(t)-\frac{f_{i+1, i}^{\ast}} {A_{i+1, i}^{\ast} } A_{i+1, i}(t). \hspace{1cm}
\end{eqnarray}
General and equilibrium solutions were obtained as follows:
\begin{eqnarray}
Y_{i+1, i}(t)=\left\{ \begin{array}{ll} 
  e^{-a(t)+a(T)}Y_{i+1, i}(T) \hspace{0.3cm} (\frac{f_{i+1, i}(t)} {A_{i+1, i}(t)}=\frac{f_{i+1, i}^{\ast}} {A_{i+1, i}^{\ast}} ),  \\\\
 e^{-a(t)+a(T)}Y_{i+1, i}(T)+e^{-a(t)} \int _{s=T}^{t}e^{a(s)}g(s) ds
\\ \hspace{0cm} (\frac{ a_{i+1, i}(t)}{dt}=A_{i+1, i}(t), g_{i+1, i}(t) = f_{i+1, i}(t)-\frac{f_{i+1, i}^{\ast}} {A_{i+1, i}^{\ast} } A_{i+1, i}(t)).
\end{array} \right.
\end{eqnarray}
%
Therefore equilibrium solutions of $X^{\ast}_{i+1, i}= \lim_{t \to \infty}X_{i+1, i}(t)$ are:
\begin{eqnarray}
\label{EquiXn}
X^{\ast}_{i+1, i}=\frac{f_{i+1, i}^{\ast}}{A_{i+1, i}^{\ast}} ,
\end{eqnarray}
where
\begin{eqnarray}
&A_{i+1, i}^{\ast}=\left\{ \begin{array}{ll} 
&2k_{2}\xi_{2}^{\ast}+2l_{2}+l_{1}  \hspace{0.5cm}(i =1), \nonumber\\ \\
&k_{2}\xi_{2}^{\ast}+k_{i+1}\xi_{i+1}^{\ast}+l_1+2l_{2}+\sum_{j= 3}^{i+ 1}l_j  \hspace{ 0.3cm}(i = 2 , \cdots,  N-1),\\\\
&k_{2}\xi_{2}^{\ast}+l_{1}+2l_{2}+\sum_{j= 3}^{n}l_j  \hspace{0.5cm}(i = N),
\end{array} \right.
\end{eqnarray}
\begin{eqnarray}
f_{i+1, i}^{\ast}=\left\{ \begin{array}{ll} 
&2k_{1}X^{\ast 2}_{1, 1}+l_2 ( \eta_{1} ^{\ast}-X^{\ast}_{1, 1}) \hspace{0.5cm}(i =1),\nonumber\\\\
&2k_{1}X^{\ast}_{1, 1}X^{\ast}_{i, i}+ k_2X_{2,1}^{\ast}X_{i-j,i}^{\ast}+ \sum_{j= 1}^{i- 1}k_{j+1}X^{\ast}_{j+1,j }X^{\ast}_{i-j,i}
+l_2 (\eta_{i}^{\ast} -\sum_{j=1}^{i}X^{\ast}_{j, i } ) 
\\& \hspace{ 0cm} +l_{i+1}( \eta_1^{\ast}-\sum_{j=1}^2 X^{\ast}_{j, 1} -\sum_{j = 1}^{i-1}X^{\ast}_{j+1, j} )
\hspace{0.5cm}(i = 2 , \cdots,  N-1),\\\\
&2k_{1}X^{\ast}_{1, 1} X^{\ast}_{N, N}+\sum_{j = 1}^{N-1} k_{j+1}X^{\ast}_{j+1,j}X^{\ast}_{N-j,N}
\hspace{ 0cm}+l_{2} \{ [b_N(0)]-\sum_{i=1}^{N} X ^{\ast}_{i, N} \}
\\& \hspace{0.3cm}(i = N).\\
\end{array}\right.
\end{eqnarray}
\subsection{Concentration of molecules in $LD_{m}$ group}
\begin{figure} [h]
  \centering
\includegraphics[width=12cm]{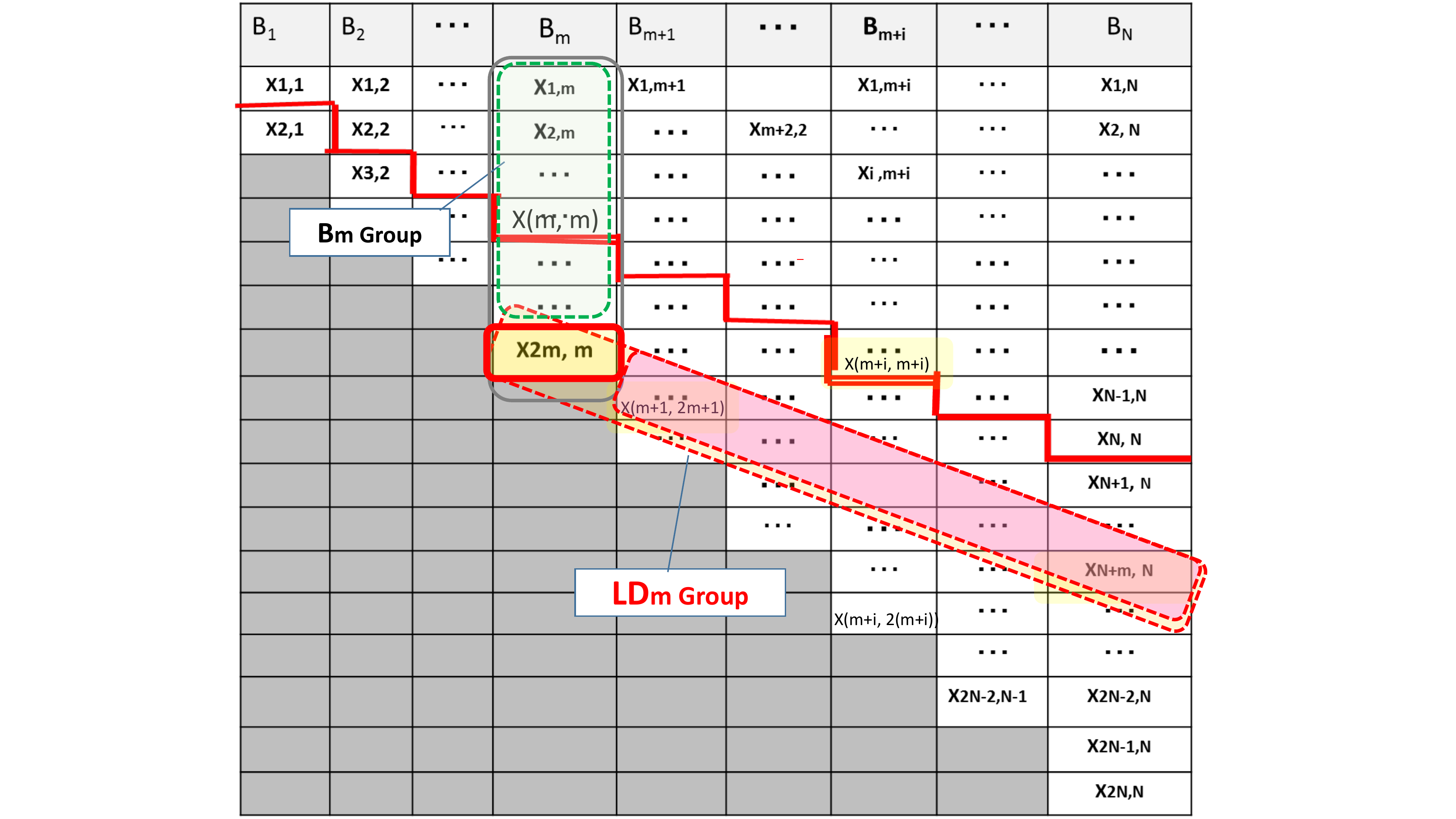}\\
  \caption{Groups related to the production of complex $X_{2m, m}\subset LD_m$ in the network diagram. Molecules in the green region of $B_m$ group react each other and produce complex $X_{2m, m}$. Molecules in the pink region of $LD_m$ group are dissolved and complex $X_{2m, m}$ is produced.  }
\label{FigLDm1}
\end{figure}

$LD_{m} (m=2,\cdots, N-1)$ group has $N-m+1$ complexes, $ X_{2m, m},  \cdots, \\ X_{2m+i, m+i}, \cdots, X_{N+m+1, N}$. 
We solved the concentrations of complex from $m= 2$ to $N-1$. A complex in $LD_m$ group can react with every complex in $LU_{m+1}$ group.
\\
From the mass conservation law and mass action law, we obtain reaction ODEs for the concentrations of $LD_{m}$ group complex, $X_{2m+i, m+i } (t) \subset B_{i}$.
\begin{eqnarray}
\label{EqXLDm}
\frac{d X_{2m+i, m+i}(t) }{dt} = -A_{2m+i, m+i}(t)X_{2m+i, m+i }(t)+f_{2m+i, m+i}(t)  \hspace{0.2 cm}(i =0 ,\cdots,  N).
\end{eqnarray}
First, we considered the concentration of the complex, $X_{2m, m} (t) \subset B_{m}, LD_{m}$ (Fig.$\ref{FigLDm1}$). This complex is symmetric, with $b_m$ monomers on both edges, $\! b_m \dots b_1b_1 \dots b_m$.  
The reactions between the complexes in  $B_N$ group lead to the generation of $X_{2m, m}$ complex. The dissolution of $LD_m$ group complexes (e.g., $X_{2m+i, m+i} (\! i= 1,\cdots,  N-m)$), leads to the generation of $X_{2m, m}$ complex and $LU_{m+1}$ group complexes (e.g., $\! X_{m+1+i, i} (\! i= 1,\cdots,  N-m-1)$).

\begin{figure} [h]
\label{LDmi}
  \centering
\includegraphics[width=12cm]{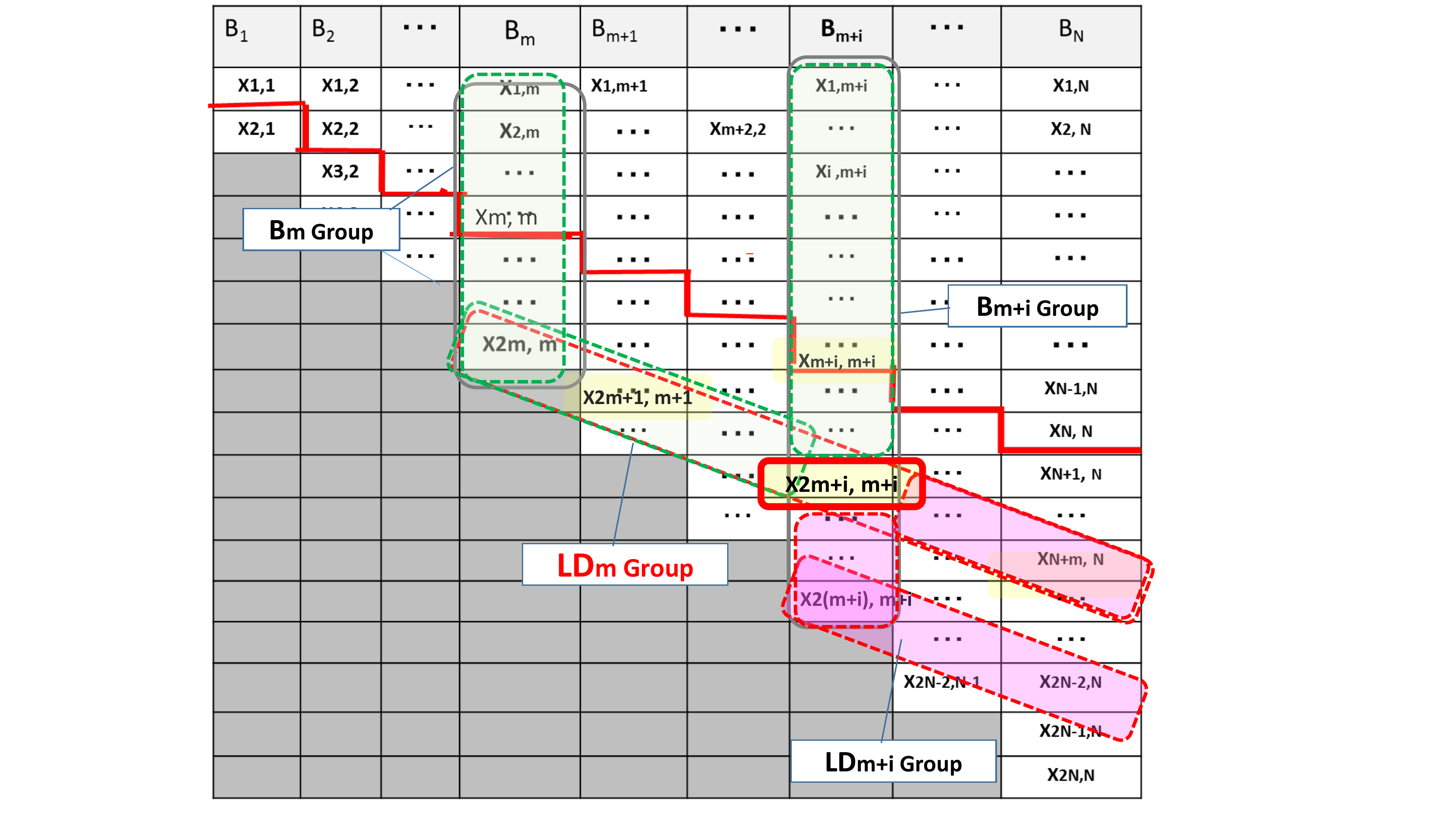}
  \caption{Groups related to the production of complex $X_{2m+i, m+i}\subset LD_m$ in the network diagram. Molecules in the green region of $B_m$ and $LD_m$ groups and molecules in the green region of $B_{m+i}$ group react each other and produce complex $X_{2m, m}$. Molecules in the pink region of $LD_m$, $B_{m+i}$ and $LD_{m+i}$ groups are dissolved and produce complex $X_{2m+i, m+i}$. }
\end{figure}

The coefficients of the equation (\ref{EqXLDm}), $A_{2m, m}$ and $ b_{2m, m}$ are defined as follows:
\begin{eqnarray}
\label{EqXLDm1Ab}
\left\{ \begin{array}{ll}
A_{2m, m }(t)&=2k_{m+1}\xi_{m+1}(t)+2\sum_{j=1}^{m+1}l_{j}  \hspace{0.3 cm}( i =0 ),\\
f_{2m, m}(t)
&=2k_{1}X_{m, m}^{2}(t) +\sum_{j=1}^{m-1}k_{j+1} X_{m-j, m }(t)X_{m+j, m}(t) 
\\&+ l_{m+1} (\eta_{m}(t) -\sum_{j=1}^{2m-1}X_{j, m}(t) )  \hspace{0.3cm}(i =0) .\\
\end{array} \right.
\end{eqnarray}
\\
Additionally, we considered the concentration of the complex, $X_{2m+i, m+i}\\ \subset B_{m+i }( \! i = 2,\cdots,  N-m-1)$ (Fig.$\ref{LDmi}$).

\begin{figure} [h]
  \centering
\includegraphics[width=12cm]{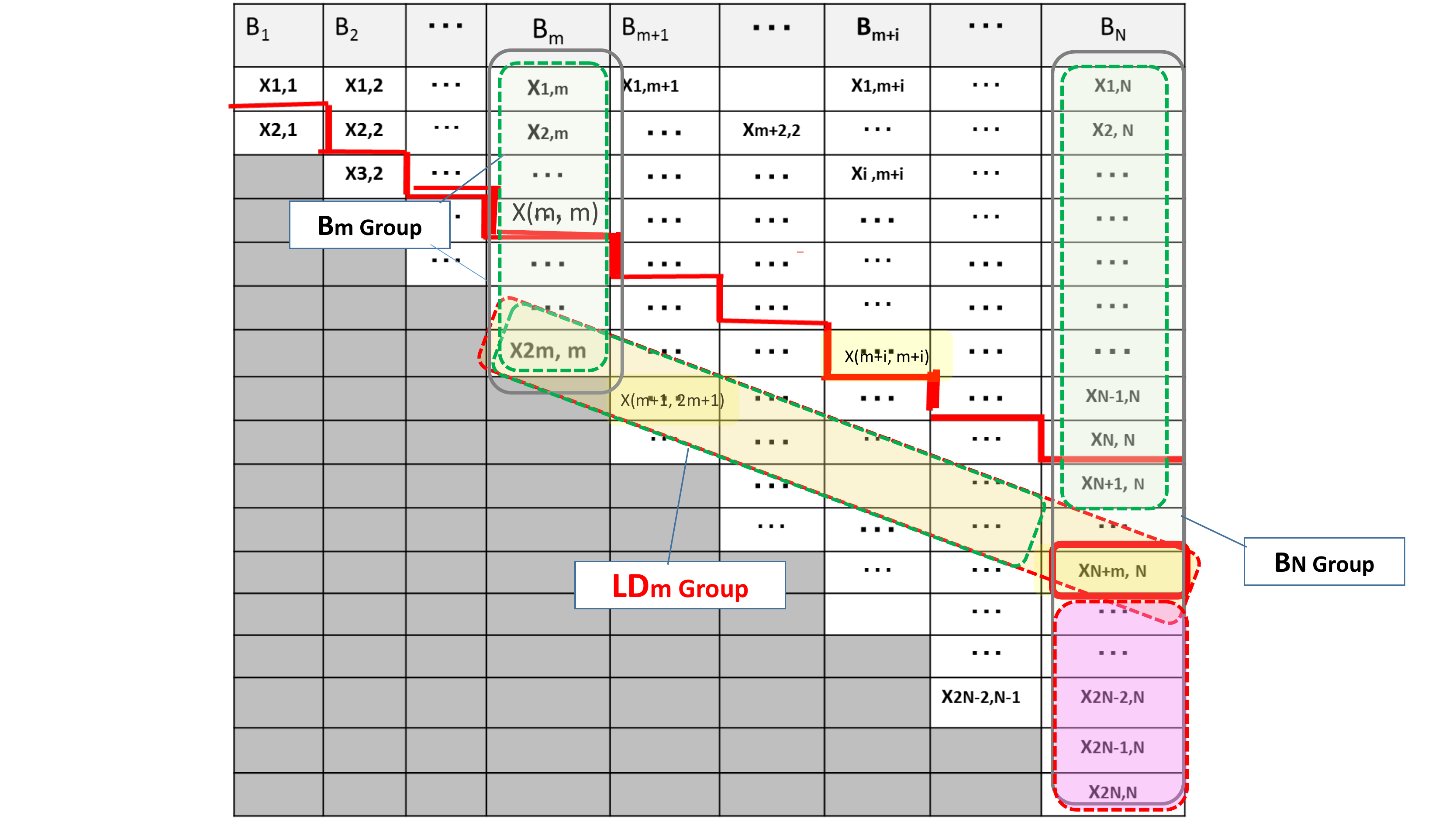}
  \caption{Groups related to the generation of complex $X_{N+m,N}\subset LD_m$ in the network diagram. Molecules in the green region of $B_m$ and $LD_m$ groups and molecules in the green region of $B_{N}$ group react each other and produce complex $X_{2m, m}$. Molecules in the pink region of $LD_N$ and $B_{N}$ groups are dissolved and produce                                         complex $X_{N+m, N}$.}
   \label{FigLDmN}
\end{figure}

\begin{eqnarray}
\label{EqXLDmiAb}
\left\{ \begin{array}{ll}
&A_{2m+i, m+i }(t)=k_{m+1}\xi_{m+1}(t)+k_{m+i+1}\xi_{m+i+1}(t)+l_{1}+2\sum_{j=2}^{m+1}l_{j}+\sum_{j=m+2}^{m+i+1}l_j , \\
&f_{2m+i, m+i}(t)=2k_1X_{m+i, m+i}(t)X_{m, m}(t)+l_{m+i}(\eta_{m+i}(t)-\sum_{j=1}^{2m+i-1}X_{j, m+i} (t))
\\&\hspace{2.5cm}+\sum_{j=2}^{m+1}k_jX_{ m+i-j+1, m+i}(t)X_{m+j-1, m} (t)
\\&\hspace{2.5cm}+\sum_{j=m+1}^{m+i}k_jX_{ m+i-j+1, m+i}(t)X_{ m+j-1, j-1}(t)
\\&\hspace{2.5cm} +\sum_{j=2}^mk_jX_{m+i+j-1,m+i }(t)X_{m-j+1, m}(t) 
\\&\hspace{2.5cm}+l_{m+2}( \eta_m(t) -\sum_{j=1}^{2m}X_{j, m}(t)-\sum_{j=0}^{i-1}X_{2m+j, m+j}(t)).
\end{array} \right.
\end{eqnarray}
I
If we consider the concentration of the complex $X_{N+m,N} (t)\subset LDm$, we observe that the $X_{N+m,N}$ complex is produced as a consequence of a reaction between a complex belonging to the $B_m$ group and a complex belonging to the $B_N$ group, or between a complex in the $LD_m$ group and another one in the $B_N$ group, and by degradation of the bigger complexes in $B_N$ group(Fig.$\ref{FigLDmN}$) 

Therefore, the coefficients of equation (\ref{EqXLDm}) were derived as follows:
\begin{eqnarray}
\label{EqXLDmnAb}
A_{N+m,N }(t)=\left\{ \begin{array}{ll}
k_{m+1}\xi_{m+1}(t)+l_{1}+2\sum_{j=2}^{m+1}l_{j}+\sum_{j=m+2}^{N}l_j (m<N+1),\\\\
k_{m+1}\xi_{m+1}(t)+l_{1}+2\sum_{j=2}^{m+1}l_{j} (m=N+1),\nonumber\\\\
\end{array} \right.
\end{eqnarray}
\begin{eqnarray}
f_{N+m,N}(t)=\left\{ \begin{array}{ll}
&\sum_{j=0}^{m-1}k_{j+1}X_{m-j, m }(t)X_{N+j, N } (t) 
+sum_{j=0}^{m}k_{j+1}X_{m+j, m}(t)X_{N-j, N } (t)\nonumber
\\&\hspace{0cm}+\sum_{j=m}^{N-1}k_{j+1}X_{m+j, j}(t)X_{N-j, N} (t) 
+ l_{m+1} ( [b_N(0)]-\sum_{j=1}^{N+m-1}X_{j, N} ).\nonumber
\end{array} \right.
\end{eqnarray}

We have already obtained the values and equilibrium solutions of the parameters $\xi_{m+1}(t), \xi_{m+i+1}(t),  X_{j, m}(t)(\!j=1,\cdots, 2m)$, $X_{2m+i, m+i}(t)(\!i=1 ,\cdots,  N-m)$.
It is possible to solve the equation (\ref{EqXLDm}), 
step by step, from $i=0$ to N, as previously done, and obtain the equilibrium solutions.

\begin{eqnarray}
\label{EquiXmi}
X^{\ast}_{2m+i, m+i} =\frac{b_{2m+i, m+i}^{\ast}}{A_{2m+i, m+i}^{\ast}} \hspace{0.5cm}( i=1 ,\cdots,  N-m),
\end{eqnarray}
where 
\begin{eqnarray}
\label{EqXLDmA}
A^{\ast}_{2m+i, m+i}=\left\{ \begin{array}{ll} 
&2k_{m+1}\xi_{m+1}^{\ast}+2\sum_{j=1}^{m+1}l_{j}  \hspace{0.5cm}(i =0),\\
&k_{m+1}\xi_{m+1}^{\ast}+k_{m+i+1}\xi_{m+i+1}^{\ast}+l_{1}+2\sum_{j=2}^{m}l_{j}+\sum_{j=m+1}^{m+i}l_j 
\\&\hspace{0.5cm}\!(i =1,\cdots,  N-m-1),\\
&k_{m+1}\xi_{m+1}^{\ast}+l_{1}+2\sum_{j=2}^{m+1}l_{j}+\sum_{j=m+2}^{N}l_j \\ &\hspace{0.5cm}(i = N-m, m<N-1),\\
&k_{m+1}\xi_{m+1}^{\ast}+l_{1}+2\sum_{j=2}^{m+1}l_{j} \hspace{0.5cm}(i = N-m, m+1 =N)\\
\end{array} \right.
\end{eqnarray}
\begin{eqnarray}
f^{\ast}_{2m+i, m+i}=\left\{ \begin{array}{ll} 
&2k_1{X^{\ast}_{m, m}}^2 +\sum_{j=1}^{m-1}k_{j+1} X^{\ast}_{m-j, m }X^{\ast}_{m+j, m}+l_{m+1}( \eta_m^{\ast}-\sum_{j=1}^{2m-1}X_{j,m}^{\ast})
\\& \hspace{0.5cm}(i =0),\\

\\& 2k_1X^{\ast}_{m+i, m+i}X^{\ast}_{m, m}
+\sum_{j=2}^{m+1}k_jX^{\ast}_{ m+i-j+1, m+i}X^{\ast} _{m+j-1, m}
\\&+\sum_{j=m+1}^{m+i}k_jX^{\ast}_{ m+i-j+1, m+i}X^{\ast}_{ m+j-1, j-1}
\\&
+\sum_{j=2}^m k_jX^{\ast}_{m+i+j-1,m+i}X^{\ast}_{m-j+1, m}
\\&+l_{m+2} ( \eta_m^{\ast} -\sum_{j=1}^{2m}X^{\ast}_{j, m}-\sum_{j=0}^{i-1}X^{\ast}_{2m+j, m+j} )
\\&+l_{m+i} ( \eta_{m+i}^{\ast}-\sum_{j=1}^{2m+i-1}X_{j, m+i } ^{\ast} )
\\&\hspace{0.5cm} (i = 2 ,\cdots,  N-m-1),\\
\\&\sum_{j=0}^{m-1}k_{j+1}X^{\ast}_{m-j, m }X^{\ast}_{N+j,N }\nonumber
 +\sum_{j=0}^{m}k_{j+1}X^{\ast} _{m+j, m }X^{\ast}_{N-j, N}
\\&+\sum_{j=m}^{N-1}k_{j+1}X^{\ast}_{m+j,j }X^{\ast}_{N-j, N }  
+ l_{m+1} ( [b_N(0)]-\sum_{j=1}^{N+m-1}X^{\ast}_{j,N} ) 
\\&(i = N-m).\\
\end{array} \right.
\end{eqnarray}
%
%
\subsection{Concentration of complexes in $LD_{N}$ group}
$LD_{N} $ group consists of only one complex, $ X_{2N, N}$. 
This complex can be produced only in the reactions between complexes in $B_N$ group, but it cannot be obtained as a product of degradation, because it represents the biggest complex in the network.
\\
 From mass conservation law and mass action law, we obtained the reaction ODE of the $X_{2N, N}$ complex as follows:
\begin{eqnarray}
\label{EqXn2n}
\frac{d X_{2N, N}(t) }{dt} = -A_{2N, N}(t)X_{2N, N}(t)+f_{2N, N}(t),  \hspace{1 cm}
\end{eqnarray}
where
\begin{eqnarray}
\label{EqXLDn2nAb}
\left\{ \begin{array}{ll}
A_{2N, N }(t)=l_{1}+2\sum_{j=2}^{N}l_{j}, \nonumber\\
f_{2N, N}(t)
=2k_{1}X^{2}_{N, N}(t) +\sum_{j = 1}^{N-1}k_{j+1}X_{j, N}(t) X_{2N-j, N}(t).\nonumber\\
\end{array} \right.
\end{eqnarray}
Coefficient  $A_{2N, N }(t)$ is constant and positive, as shown in the equation $(\ref{EqXLDn2nAb})$.
We set this coefficient as $A_{2N, N}^{\ast}=l_{1}+2\sum_{j=2}^{N}l_{j}$, which allowed us to solve the equation $(\ref{EqXn2n})$ by varying the parameters, and $X_{2N, N}(0)=0$.
\begin{eqnarray}
\label{SolXn2n}
X_{2N, N}(t) = \int_0^t e^{-(t-s)A_{2N, N}^{\ast}} f_{2N, N}(s) ds.  
\end{eqnarray}

When  the parameter $Y_{2N, N}(t)$ is set as $Y_{2N, N}(t)=X_{2N, N}(t)-\frac{f_{2N, N}^{\ast}}{A_{2N, N}^{\ast}}$, from the equation $(\ref{EqXn2n})$, we can obtain:
\begin{eqnarray}
\label{EqYn2n}
\frac{d Y_{2N, N}(t) }{dt} = -A_{2N, N}^{\ast}Y_{2N, N}(t)+f_{2N, N}(t) - f_{2N, N}^{\ast}. \hspace{1 cm}
\end{eqnarray}
%
We obtained the equilibrium solutions as previously:
\begin{eqnarray}
\label{EquiXn2n}
X^{\ast}_{2N, N} =\frac{f_{2N, N}^{\ast}}{A_{2N, N}^{\ast}},\hspace{0.5cm}( \!i=1,\cdots, N-m),
\end{eqnarray}
where 
\begin{eqnarray}
\left\{ \begin{array}{ll}
A_{2N, N}^{\ast}=l_{1}+2\sum_{j=2}^{N}l_{j}\nonumber \\
f_{2N, N}^{\ast}=2k_{1}X^{\ast 2}_{N, N} +\sum_{j = 1}^{n-1}k_{j+1}X^{\ast}_{j, N}X^{\ast}_{2N-j, N}. \nonumber
\end{array} \right.
\end{eqnarray}
This led to a conclusion that all complexes in the $(b_N-\cdots -b_1-b_1- \cdots-b_N) $ type network  are integrable and have explicit equilibrium solutions. 

\section{Simulation }
We applied the equilibrium values of complex concentrations, determined for the N monomer network, to the 3-monomer MMP2/TIMP2/MT1-MMP network. The equilibrium values of the group and complex concentrations of this network were calculated using R  ver. 3.2.2 wuth desolve package. 

The parameter $X_{4,3}(t)$ is a concentration of the complex (MMP2/TIMP2/MT1-MMP/MT1-MMP) at time $t$, and $b_1$ corresponds to MT1-MMP, $b_2$ to TIMP2, $b_3$ to MMP2.

\begin{figure}
  \begin{center}
\includegraphics[width=.80\linewidth]{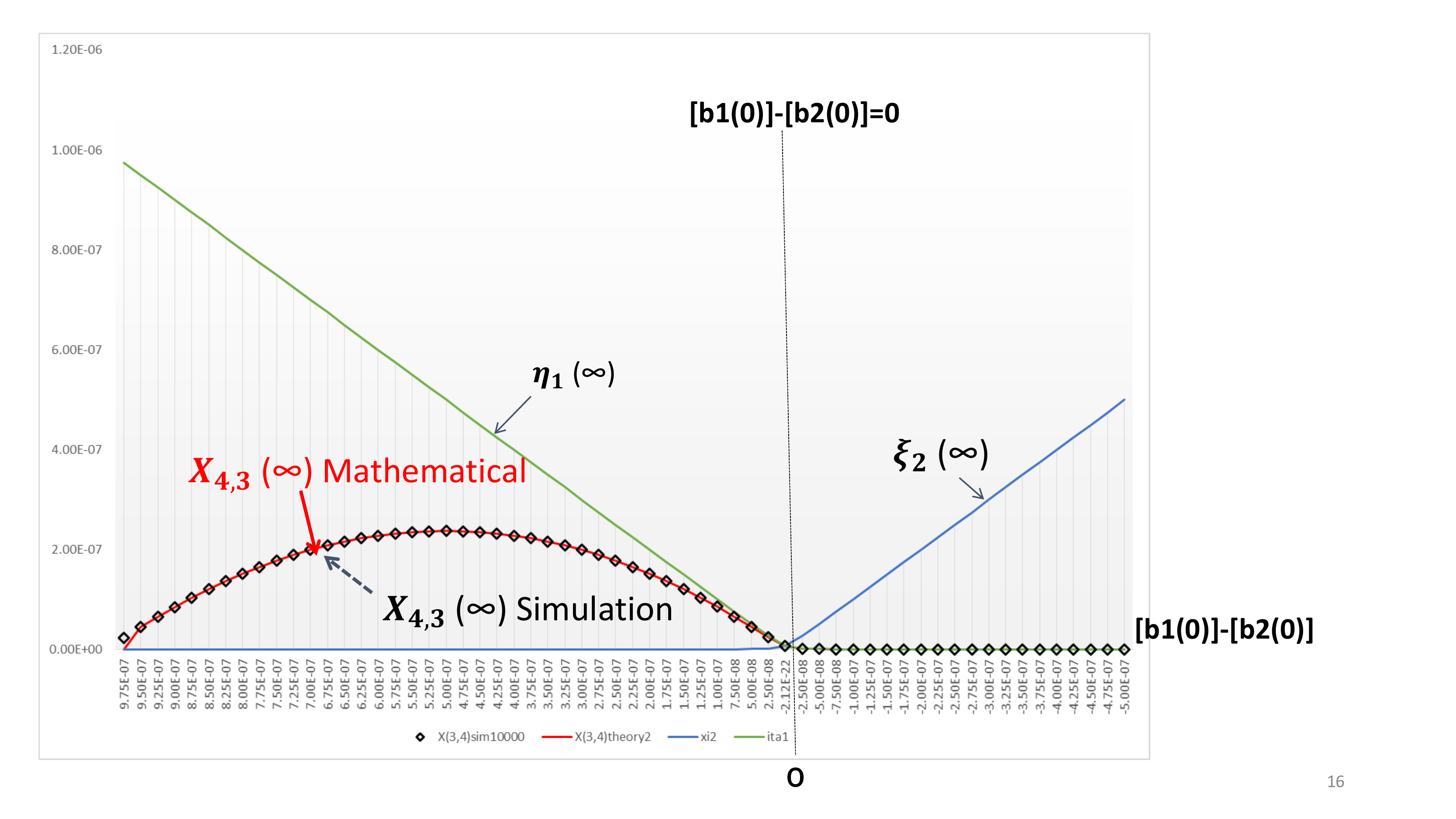}
\end{center}
  \caption{Simulation and theoretical results of the equilibrium concentrations of the molecule $X_{4,3}$,  $\eta_1$ group, and $\xi_2$ group at $[b_1(0)]=1000nM$, during the MMP2 activation process.
      $X_{4,3}$ corresponds to the (MMP2/TIMP2/MT1-MMP/MT1-MMP) complex. The black dotted line represents $X_{4,3}(\infty)$, according to ODE simulations. Red line represents the concentration of $X_{4,3}(\infty)$ molecule, calculated using the equation(68). The results we obtained and the simulation result show a high level of agreement. Green line represents the equilibrium group concentration $\eta_1(\infty)$, where the theoretical equations were applied. $\eta_1$ group includes the target molecule $X_{4,3}$. Blue line represents the equilibrium group concentration $\xi_2(\infty)$, where we applied our theoretical equations. $\xi_2$ group has dual relationship with $\eta_1$ group. The positive or negative difference in the concentrations, $[b_1(0)]-[b_2(0)]$, determines the positive and zero $\eta_1(\infty)$ and $\xi_2(\infty)$ groups.}
\label{2dsim}
\end{figure}

From this simulation, we can see that the theoretical result of the target complex concentration $X_{4,3}(\infty)$ agrees with the results obtained by the ODE simulations of the target complex concentration $X_{4,3}(\infty)$ (Fig.\ref{2dsim}). The group concentration $\eta_1(\infty)$ was shown to be always bigger than $X_{4,3}(\infty)$.

The difference between monomers in the network, $[b_2(0)]-[b_1(0)]$, represents a regulatory parameter for the $\eta_1$ group, $\xi_2$ group, and the complexes that belong to these groups, such as $X_{4,3}$. When $[b_2(0)]-[b_1(0)]<0$, $\eta_1(\infty)$ and $X_{4,3}(\infty)$ are positive and $\xi_2(\infty) =0$. When $[b_2(0)]-[b_1(0)]=0$, $\eta_1(\infty)$, $\xi_2(\infty)$, and $X_{4,3}(\infty)$ are 0. When $[b_2(0)]-[b_1(0)]>0$, $\eta_1(\infty)=X_{4,3}(\infty)=0$, and $\xi_2(\infty)$ is positive. 

This demonstrates that this is crucial for the determination of $\eta_1(\infty)$, $\xi_2(\infty)$, and $X_{4,3}(\infty)$ concentrations.

\begin{figure}
  \begin{center}
\includegraphics[ width=.85\linewidth]{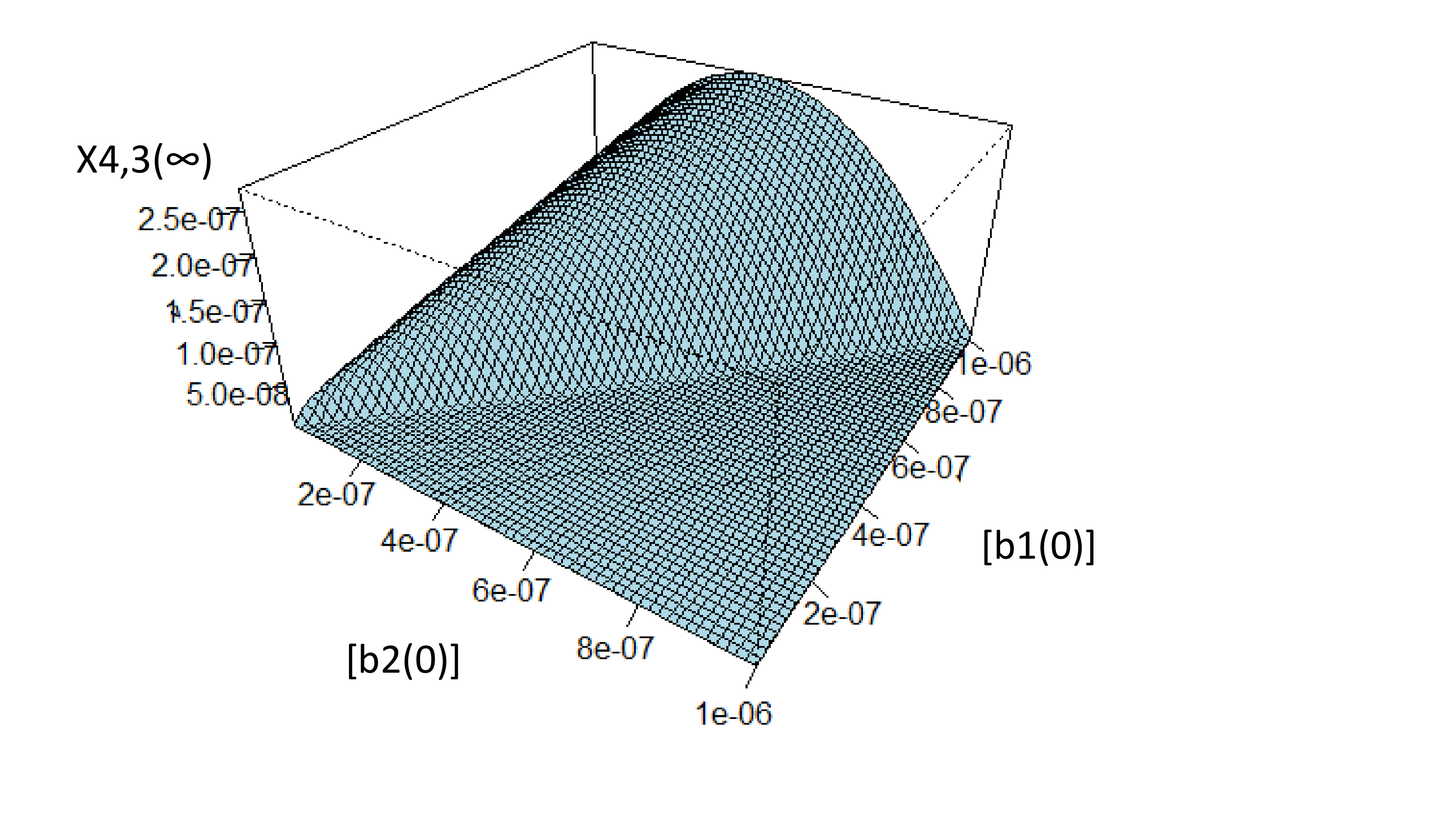}
 \end{center}
 \caption{The result of the equilibrium concentration simulation for the molecule $X_{4,3}$ when $[b_1(0)]$ and $[b_2(0)]$ vary from 0 nM to 1000 nM during the MMP2 activation process.}
\label{3dsim}
\end{figure}

Figure \ref{3dsim} shows  $X_{4,3}(\infty)$ when $[b_1(0)]$ and $[b_2(0)]$ vary from 0 to 1000 nM.
For $[b_2(0)]-[b_1(0)]>0$, $X_{4,3}(\infty)$ is always 0.

%
\section{Conclusions}
By generalizing MMP2 activation network to N monomer network with $(b_N\cdots b_1-b_1 \cdots b_N) $ complexes, we can observe the group reaction behavior and the layered structure of the network. This allowed us to take a two-step approach, in order to solve the pathway network system.

As the first step, we classified the complexes to the reaction groups, such as $B_m, LU_m$, and  $LD_m$, and these groups behave according to the mass conservation laws. Therefore, we aggregated complex ODEs into the reaction group ODEs, and obtained the strict solutions for reaction group concentrations. In the following step, we solved all complex concentrations in the each reaction group, while the group concentrations were obtained at the first step. 
The concentrations of all complexes in the N monomer pathway network are derived explicitly  and converge to equilibria.
 
We think our approach is useful for biologists dealing with the pathway network with homodimer symmetric formation complexes. We show this type pathway network is integrable and stable. 
Each concentration of the complexes in the network is solved explicitly.
Especially, we obtain equilibrium concentration of each complex  easily and quickly with our approach.

We also identify the difference between the initial concentrations of the reacting monomers, $[b_m(0)]-[b_{m+1}(0)]$, as regulator parameter of the target complex behavior.  
For example, in the case of MMP2 activation pathway network, the difference of the initial concentration between monomer $b_1$:MT1-MMP and monomer $b_2$:TIMP2, $[b_1(0)]-[b_{2}(0)]$ is regulator of the target complex $X(4,3)$, as shown at simulation section. This parameter decides the behavior and equilibrium of target complex or the groups it belongs. 

When not all the reaction rate constants and initial concentrations but the relevant constants and initial concentrations are known,  we can calculate the group concentrations the target complex related, as the upper value of the target complex.  
It is important to obtain the upper value of the target complex concentration in the difficult case to get all informations of the pathway network. 
It is difficult to simulate the target complex concentration  directly with the ODEs. 



Our mathematical approach may help us understand the mechanism of this type pathway network by knowing the background mathematical laws which govern the network.
We think that pathway networks and chemical networks with homodimer symmetric complex formation  $(b_N\cdots b_1b_1\cdots b_N) $ have similar dominant mathematical laws. Autophagy complex and its effects on the production of Atg-PE8 are very similar to  MMP2 activation and (MT1-MMP/TIMP2/MMP2) complex formation.
In this research, we take account of only chemical reactions of monomers and complexes for simplicity.

\section*{Competing interests}
The authors declare that they have no competing interests.

\section*{Author's contributions}
 The original mathematical models were drawn from earlier work of TS. Modifications were made by KI.
Modeling and computational works were carried out by the first author with suggestions from the second.
All other works  were executed jointly.

\section*{Acknowledgements}
This work is founded by JSPS Core-to-Core Program ``Establishing International Research Network of Mathematical Oncology''.

\appendix
\section{Group Concentration}
\subsection{$LU_{m}$ group}
As we saw previously, the reaction groups follow mass preservation laws.
$LU_{m+1}$ group has dual relationships with $B_m$ and $LD_m$ groups, as seen in equation $(\ref{EqMCLnh})$ ($m =1, \cdots,  N-1$).

We sum ODEs of the molecules in the $LU_{m+1}$ group as:
\begin{eqnarray}
	&&\frac{d\sum_{total}LU_{m+1}(t)}{dt} \nonumber\\
	&=& \sum_{i=1}^{N-m} \frac{d X_{i, m+i }(t)}{dt} \nonumber\\
	&=&-k_{m+1}\sum_{total}LU_{m+1}(t)\{ \sum_{total}B_{m}(t)+\sum_{total}LD_{m}(t) \} -l_{m+1}\sum_{total}LU_{m+1}(t)\nonumber\\
     && +l_{m+1} \{\sum_{j=1}^{N-m}( \sum_{total}B_{m+j}(t)+\sum_{total}LD_{m+j} (t))-\sum_{j=2}^{N-m}\sum_{total}LU_{m+j}(t) \}.\nonumber
\end{eqnarray}
%
We define the parameter $\xi_ {m}(t)$ as the total concentration of the molecules belonging to the $LU_{m}$ group.
The parameter $\eta_{m}(t)$ is defined as the total concentration of molecules in $B_{m}$ and $LD_{m}$ groups.

\begin{eqnarray}
\left\{
\begin{array}{l}
\xi_ {m}(t)= \Sigma _{total}LU_{m}(t), \\
\eta_{m}(t)=\Sigma _{total}(B_m(t)+LD_m(t)).
\end{array}
\right.
\end{eqnarray}

Afterward, equations ($\ref{EqMCLnh}$)-($\ref{EqMCLh1}$) are rewritten using $\xi_{m+1}(t)$ and $\eta_m(t)$:
\begin{eqnarray}
\label{EqMCLm}
[b_{m}(0)]-[b_{m+1}(0)]= \eta_m(t) -\xi_{m+1}(t).  
\end{eqnarray}
\begin{eqnarray}
[b_{m+1}(0)]=\Sigma _{j=m+1}^{N} \eta_j (t)-\Sigma _{j=m+1}^{N-1} \xi_{j+1}(t).
\end{eqnarray}

Then,
\begin{eqnarray}
\label{EqXim}
\frac{d \xi_{m+1}(t)}{dt} 
= -k_{m+1}\xi_{m+1}(t)^2-\{ ( [b_{m}(0)]-[b_{m+1}(0)])k_{m+1}+l_{m+1} \} \xi_{m+1}(t)\nonumber\\
+l_{m+1}[b_{m+1}(0)] .
\end{eqnarray}

We obtain $LU$ group ODEs in this way ($m=1,\cdots, N$).
$N(N+1)$ molecule ODEs were aggregated to $N$ reaction group ODEs, which can be solved explicitly and have asymptotically stable solutions.  \\
The discriminant of $\frac{d \xi_{m+1}(t)}{dt}$ is:
\begin{eqnarray}
D_{m+1}= \{k_{m+1}( [b_{m}(0)]-[b_{m+1}(0)])+l_{m+1}\}^2 +4k_{m+1} l_{m+1} [b_{m+1}(0)].
\end{eqnarray}

If $D_{m+1}$ is positive, there are two equilibrium solutions of $\frac{d \xi_{m+1}(t)}{dt}$ = 0.
\begin{eqnarray}
\xi_{m+1 \pm}^{\ast}=\frac{- \{ k_{m+1}([b_{m}(0)]-[b_{m+1}(0)])+l_{m+1}\} \pm\sqrt {\mathstrut D_{m+1} }} {4k_{m+1}}.
\end{eqnarray}
The explicit solution of equation($\ref{EqXim}$) is:
\begin{eqnarray}
\xi_{m+1}(t)=\frac{\xi_{{m+1} +}^{\ast}-C_{m+1}\xi_{{m+1} -}^{\ast}e^{-\beta_{m+1}t}}{1-C_{m+1}e^{-\beta_{m+1}t}},
\end{eqnarray}
where $C_{m+1}=\frac{[b_{m+1} (0)]+\xi_{m+1  -}^*}{[b_{m+1} (0)] +\xi_{m+1 +}^*}, \beta_{m+1} =k_{m+1}( {\xi_{{m+1} +}^*-\xi_{{m+1} -}^*})$.\\

When time $t$ tends to $+{\infty}, e^{-\beta_{m+1}t} \rightarrow +0$.
\begin{eqnarray}
\label{EqEquiXi}
\lim_{t\rightarrow\infty}\xi_{m+1}(t)=\xi_{{m+1} +}^{\ast}.
\end{eqnarray}

If  $l_{m+1}=0$ and $[b_{m}(0)]=[b_{m+1}(0)]$, the discriminant $D_{m+1}=0$.  
In this case, we obtain a steady solution of equation($\ref{EqXim}$) as $\xi_{m+1}^{\ast}=0$. 
The equation($\ref{EqXim}$) is rewritten as follows:
\begin{eqnarray}
\frac{d \xi_{m+1}(t)}{dt}=-k_{m+1}  {\xi_{m+1}(t)}^2.
\end{eqnarray}
The explicit solution of equation($\ref{EqXim}$) in this case is:
\begin{eqnarray}
{\xi_{m+1}(t)}=  \frac{[b_{m+1}(0)]}{k_{m+1}[b_{m+1}(0)]t+1}.
\end{eqnarray}
${\xi_{{m+1}}(t)}$ decreases with $O(t^{-1})$ more slowly than in the case where $D_{m+1} >0$ with $O(e^{-\beta_{m+1}t})$.
When time $t$ tends to +$\infty$, the numerator is constant and denomination tends to +$\infty$:
\begin{eqnarray}
\lim_{t\rightarrow\infty}\xi_{m+1}(t)=\xi_{m+1}^{\ast}=0.\nonumber
\end{eqnarray}
After obtaining $LU_{m+1}$ group concentration $\xi_{m+1}(t)$ and its equilibrium value $\xi_{m+1}^{\ast}$, we obtained the $B_m$ group and $LD_m$ group concentration $\eta_{m}(t)$ and its equilibrium value $\eta_{m}^{\ast}$, using the mass conservation law equation ($\ref{EqMCLm}$). 
\begin{eqnarray}
	\eta_{m}(t) =[b_{m}(0)]-[b_{m+1} (0)] +\xi_{m+1} (t).
\end{eqnarray}
\begin{eqnarray}
	\eta_{m}^{\ast} =[b_{m}(0)]-[b_{m+1} (0)] +\xi_{m+1} ^{\ast}.
\end{eqnarray}

$LU_N$ and $LU_1$ reaction group ODEs are somewhat different from $LU_m$ reaction groups ($m=2 \cdots, N-1$).

\subsection{$LU_{N}$ Group}
$LU_N$ group consists of only molecule $X(1,N)$, and the ODE of the molecule $X(1,N)$ is:
\begin{eqnarray}
\frac{d X_{1,N}(t)}{dt}= -k_{N}X_{1,N}(t) ( \sum_{total}B_{N-1}+\sum_{total}LD_{N-1} ) 
\nonumber\\
\hspace{1.5cm}+l_{N}(\sum_{total}B_{N}+\sum_{total}LD_{N}- X_{1,N}(t)).
\end{eqnarray}

The $LU_N(t)$ reaction group concentration is equal to the molecule concentration $X_{1,N}(t)$.
Therefore, the ODE of the $LU_N$ reaction group is:
 \begin{eqnarray}
\label{EqXiN}
\frac{d \xi_N (t)}{dt}&=& -k_{N}\xi_N(t) ([b_{N-1}(0)]-[b_N(0)] +\xi_N(t)) +l_N([b_N(0)]-\xi_N(t)) \nonumber\\
&=&-k_{N}\xi_N(t)^2 -\{k_{N} ([b_{N-1}(0)]-[b_N(0)])+ l_N \}\xi_N(t) +l_N [b_N(0)]. \nonumber\\
\end{eqnarray}
The solutions for the equation ($\ref{EqXiN}$) are obtained in the similar way as $LU_m$ reaction group concentration.
The discriminant of $\frac{d \xi_{N}(t)}{dt}$ is:
\begin{eqnarray}
D_{N}= \{k_{N}( [b_{N-1}(0)]-[b_{N}(0)])+l_{N}\}^2 +4k_{N} l_{m+1} [b_{N}(0)].
\end{eqnarray}
If $D_{N}$ is positive, there are two equilibrium solutions of $\frac{d \xi_{N}(t)}{dt}= 0$.
\begin{eqnarray}
\xi_{N \pm}^{\ast}=\frac{- \{ k_{N}([b_{N-1}(0)]-[b_{N}(0)])+l_{N}\} \pm\sqrt {\mathstrut D_{N} }} {4k_{N}}.
\end{eqnarray}
The explicit solution of equation ($\ref{EqXiN}$) is:
\begin{eqnarray}
\xi_{N}(t)=\frac{\xi_{{N} +}^{\ast}- C_{N}\xi_{{N} -}^{\ast}e^{-\beta_{N}t}}{1-C_{N}e^{-\beta_{N}t}}.
\end{eqnarray}
where $C_{N}=\frac{[b_{N} (0)]+\xi_{N  -}^*}{[b_{N} (0)] +\xi_{N +}^*}, \beta_{N} =k_{N}( {\xi_{{N} +}^*-\xi_{{N} -}^*})$.\\
When time $t$ tends to $+{\infty}, e^{-\beta_{N}t} \rightarrow +0$:
\begin{eqnarray}
\label{EqEquiXiN}
\lim_{t\rightarrow\infty}\xi_{N}(t)=\xi_{{N} +}^{\ast}.
\end{eqnarray}
In the case when $l_{N}=0$ and $[b_{N-1}(0)]=[b_{N}(0)]$, i.e., the discriminant $D_{N}=0$,
the equation ($\ref{EqXiN}$) is rewritten as follows:
\begin{eqnarray}
\frac{d \xi_{N}(t)}{dt}=-k_{N}  {\xi_{N}(t)}^2.
\end{eqnarray}
The explicit solution of equation ($\ref{EqXiN}$) in this case is:
\begin{eqnarray}
{\xi_{N}(t)}=  \frac{[b_{N}(0)]}{k_{N}[b_{N}(0)]t+1}.
\end{eqnarray}
Therefore, the equilibrium solution of equation ($\ref{EqXiN}$) is:
\begin{eqnarray}
\lim_{t\rightarrow\infty}\xi_{N}(t)=\xi_{N}^{\ast}=0.\nonumber
\end{eqnarray}
We obtained $B_{N-1}$ and $LD_{N-1}$ group concentration $\eta_{N-1}(t)$ and the equilibrium value $\eta_{N-1}^{\ast}$, using the mass conservation law equation ($\ref{EqMCLm}$). 
\begin{eqnarray}
	\eta_{N-1}(t) =[b_{N-1}(0)]-[b_{N} (0)] +\xi_{N} (t).
\end{eqnarray}
\begin{eqnarray}
	\eta_{N-1}^{\ast} =[b_{N}(0)]-[b_{N} (0)] +\xi_{N} ^{\ast}.
\end{eqnarray}
%
%
\subsection{$LU_{1}$ reaction group}
Next, we considered {$LU_{1}$ reaction group.  
{$LU_{1}$ reaction group consisted of monomer $X_{1,1}$, $X_{2,2},\cdots, X_{i, i} \cdots$ complexes, and $X_{N,N}$. 
$\frac{d \xi_{1}(t)}{dt}$ is obtained as the sum of the upper ODEs. 
The reaction terms defining which monomers or complexes of group $LU_{1}$ lead to the formation of other monomers or complexes of this group are canceled.
\begin{eqnarray}
	\frac{d \xi_{1}(t)}{dt} 
					= -2k_{1}\xi_{1}(t)^2+l_{1} \{ [b_1(0)]-\xi_{1}(t) \}\nonumber\\
= -2k_{1}\xi_{1}(t)^2-l_{1}\xi_{1}(t)+l_{1}  [b_1(0)] .
\end{eqnarray}
The discriminant of $\frac{d \xi_{1}(t)}{dt}$ is:
\begin{eqnarray}
D_{1}=l_{1}^2 +8k_1 l_{1} [b_1(0)].
\end{eqnarray}
If $D_1$ is greater than 0, i.e., $l_1>0$, there are two equilibrium value solutions of $\frac{d \xi_{1}(t)}{dt}=0$:
\begin{eqnarray}
\xi_{1 \pm}^{\ast}=\frac{- l_{1} \pm\sqrt {\mathstrut l_{1} ^2+8k_1 l_{1} [b_1(0)] }}{4k_1}.
\end{eqnarray}
We considered $\xi_1(t)$ as the total monomer and complex concentrations in group $LU_1$:
\begin{eqnarray}
\xi_{1}(t)=\frac{\xi_{1 +}^{\ast}- C_{1}\xi_{1 -}^{\ast}e^{-\beta_{1}t}}{1-C_1e^{-\beta_{1}t}},
\end{eqnarray}
where $C_{1}=\frac{[b_{1} (0)]+\xi_{1 -}^*}{[b_{1} (0)] +\xi_{1 +}^*}, \beta =k_{1}( {\xi_{1 +}^*-\xi_{1 -}^*})$.
When time t tends to $+{\infty}, e^{-\beta_{1}t} \rightarrow +0$:
\begin{eqnarray}
\lim_{t\rightarrow\infty}\xi_1(t)=\xi_{1 +}^{\ast}.\nonumber
\end{eqnarray}
The discriminant $D_1=0$ when  $l_{1}=0$.  In this case, we obtained a steady solution for the equation (3.1), $\xi_{1 }^{\ast}=0$ and equation $\frac{d \xi_{1}(t)}{dt}=-2k_1  {\xi_{1}(t)}^2 $ from (3.1). 
The total monomer and complex concentrations in group $LU_1$ were:
\begin{eqnarray}
{\xi_{1}(t)}=  \frac{[b_1(0)]}{k_1[b_1(0)]t+1},\nonumber 
\end{eqnarray}
and ${\xi_{1}(t)}$ decreases with $ O(t^{-1})$ more slowly than in the case $D_1 >0$ with $O(e^{-\beta_{1}t})$.
When time t tends to +$\infty$, the numerator is constant and denominator tends to +$\infty$.
\begin{eqnarray}
\lim_{t\rightarrow\infty}\xi_1(t)=\xi_{1}^{\ast}=0.\nonumber
\end{eqnarray}
We obtain $B_{1}$ and $LD_{1}$ group concentration $\eta_{1}(t)$, and the equilibrium value $\eta_{1}^{\ast}$, through the mass conservation law equation ($\ref{EqMCLm}$):
\begin{eqnarray}
	\eta_{1}(t) =[b_{1}(0)]-[b_{2} (0)] +\xi_{1} (t).
\end{eqnarray}
\begin{eqnarray}
	\eta_{1}^{\ast} =[b_{1}(0)]-[b_{2} (0)] +\xi_{1} ^{\ast}.
\end{eqnarray}
The LU reaction group concentrations are summarized as follows:
\begin{eqnarray}
\xi_{N}(t)&=\left\{ \begin{array}{ll} 
\frac{\xi_{{N} +}^{\ast}- C_{N}\xi_{{n} -}^{\ast}e^{-\beta_{n}t}}{1-C_{n}e^{-\beta_{n}t}},& (l_{N}>0, m=N),\\
\frac{[b_{N}(0)]}{k_{N}[b_{N}(0)]t+1}, & (l_{N}=0$ and $[b_{N-1}(0)]=[b_{N}(0)], m=N).\\
\end{array} \right.
\end{eqnarray}

\begin{eqnarray}
\xi_{m}(t)&=\left\{ \begin{array}{ll} 
\frac{\xi_{{m} +}^{\ast}- C_{m}\xi_{{m} -}^{\ast}e^{-\beta_{m}t}}{1-C_{m}e^{-\beta_{m}t}},& (l_{m}>0, m:1\downarrow N-1),\\
\frac{[b_{m}(0)]}{k_{m}[b_{m}(0)]t+1},& (l_{m}=0$ and $[b_{m-1}(0)]=[b_{m}(0)],m:1\downarrow N-1).\\
\end{array} \right.
\end{eqnarray}

\begin{eqnarray}
\xi_{1}(t)&=\left\{ \begin{array}{ll} 
\frac{\xi_{1 +}^{\ast}- C_{1}\xi_{1 -}^{\ast}e^{-\beta_{1}t}}{1-C_1e^{-\beta_{1}t}},&(l_{1}>0, m=1),\\
\frac{[b_1(0)]}{k_1[b_1(0)]t+1},& (l_{1}=0, m=1).\\
\end{array} \right.
\end{eqnarray}
LU group concentrations, $\xi_{N},...,\xi_{m},...,\xi_{1}$, are independent of each other and have strict solutions, which converge to equilibrium solutions when time t tends to $+\infty$.\\
The total concentration of $B_m$ and $LD_m$ groups $\eta_m$ is derived from the equation ($\ref{EqMCLnh}$), as follows:
\begin{eqnarray}
\left\{\begin{array}{ll} 
\eta_{N}(t)&=[b_N(0)], \hspace{1cm}(m=N),\\
\eta_{m}(t)&=[b_{m}(0)]-[b_{m+1}(0)]+\xi_{m+1}(t), \hspace{1cm}(m=1,\cdots, N-1).
\end{array} \right.
\end{eqnarray}
Here, we showed that the total concentration of $LU_m$ group, or total concentrations of $LD_m$ and $B_m$ groups, can be solved. They converge to the equilibrium solutions when time t tends to $+\infty$.

We regarded the reaction group concentrations as a priori upper boundaries of the concentration of each molecular in the group.
Afterward, we demonstrated that each molecule in a group converges to a stable solution.

\end{document}